\documentclass[prr,superscriptaddress,floatfix,twocolumn,showpacs]{revtex4-2}
\usepackage{babel}
\usepackage{fmtcount}
\usepackage{verbatim}
\usepackage{amsmath}
\usepackage{amsfonts}
\usepackage{amssymb}
\usepackage{graphicx}
\usepackage{bm}
\usepackage{color}
\usepackage{hyperref}
\usepackage[active]{srcltx}
\usepackage{cases}
\usepackage{ulem}
\usepackage{multirow}
\usepackage{braket}
\usepackage[utf8]{inputenc}

\begin{document}

\title{Distinguishability Transitions in Non-Unitary Boson Sampling Dynamics}
\author{Ken Mochizuki}
\affiliation{Nonequilibrium Quantum Statistical Mechanics RIKEN Hakubi Research Team, RIKEN Cluster for Pioneering Research (CPR), 2-1 Hirosawa, Wako 351-0198, Japan}

\author{Ryusuke Hamazaki}
\affiliation{Nonequilibrium Quantum Statistical Mechanics RIKEN Hakubi Research Team, RIKEN Cluster for Pioneering Research (CPR), 2-1 Hirosawa, Wako 351-0198, Japan}
\affiliation{RIKEN Interdisciplinary Theoretical and Mathematical Sciences Program (iTHEMS), 2-1 Hirosawa, Wako 351-0198, Japan}

\begin{abstract}
We discover novel transitions characterized by distinguishability of bosons in non-unitary dynamics with parity-time ($\mathcal{PT}$) symmetry. 
We show that $\mathcal{PT}$ symmetry breaking, a unique transition in non-Hermitian open systems, enhances regions in which bosons can be regarded as distinguishable. 
This means that classical computers can sample the boson distributions efficiently in these regions by sampling the distribution of distinguishable particles. 
In a $\mathcal{PT}$-symmetric phase, we find one dynamical transition upon which  the distribution of bosons deviates from that of distinguishable particles, when bosons are initially put at distant sites. 
If the system enters a $\mathcal{PT}$-broken phase, the threshold time for the transition is suddenly prolonged, since dynamics of each boson is diffusive (ballistic) in the $\mathcal{PT}$-broken ($\mathcal{PT}$-symmetric) phase. 
Furthermore, the $\mathcal{PT}$-broken phase also exhibits a notable dynamical transition on a longer time scale, at which the bosons again become distinguishable. 
This transition, and hence the classical easiness of sampling bosons in long times, are true for generic postselected non-unitary quantum dynamics, while it is absent in unitary dynamics of isolated quantum systems. 
$\mathcal{PT}$ symmetry breaking can also be characterized by the efficiency of a classical algorithm based on the rank of matrices, which can (cannot) efficiently compute the photon distribution in the long-time regime of the $\mathcal{PT}$-broken ($\mathcal{PT}$-symmetric) phase.
\end{abstract}

\maketitle

\section{Introduction}
\label{sec:introduction}
The last decade has witnessed a rapid progress of our understanding in non-Hermitian quantum mechanics, which provides an effective description of open quantum systems that exhibit non-unitary dynamics. 
Indeed, various intriguing phenomena have been revealed, such as effects of disorder \cite{hatano1996localization,
shnerb1998winding,
feinberg1999non,
kalish2012light,
basiri2014light,
mochizuki2017effects,
hamazaki2019non,
hamazaki2020universality,
mochizuki2020statistical,
tang2020topological,
tzortzakakis2020non,
luo2021universality}, 
parity-time ($\mathcal{PT}$) symmetry breaking \cite{bender1998real,
bender2002complex,
bender2002generalized,
bender2003must,
mostafazadeh2003exact,
bender2007making,
guo2009observation,
ruter2010observation,
lin2011unidirectional,
chtchelkatchev2012stimulation,
miri2012large,
regensburger2012parity,
peng2014parity,
feng2014single,
hodaei2014parity,
mochizuki2016explicit,
ashida2017parity,
xiao2017observation,
longhi2018parity,
el2018non,
li2019observation,
longhi2019non,
kawasaki2020bulk,
mochizuki2020bulk,
takasu2020pt}, and topological phenomena \cite{esaki2011edge,
hu2011absence,
poli2015selective,
lee2016anomalous,
xiao2017observation,
bandres2018topological,
chen2018characterization,
gong2018topological,
harari2018topological,
kunst2018biorthogonal,
parto2018edge,
xiao2018higher,
yao2018edge,
lee2019anatomy,
sone2019anomalous,
song2019non,
yokomizo2019non,
kawabata2019classification,
kawabata2019symmetry,
helbig2020generalized,
mochizuki2020bulk,
okuma2020topological,
sone2020exceptional,
xiao2020non,
weidemann2020topological,
haga2021liouvillian,
pan2021point}. 
Interestingly, non-Hermitian dynamics has been observed not only in genuinely quantum systems, such as photonic systems  \cite{xiao2017observation,xiao2020non,tang2016experimental} and ultracold atomic systems \cite{li2019observation,takasu2020pt},  but also in classical systems \cite{guo2009observation,
ruter2010observation,
regensburger2012parity,
peng2014parity,
feng2014single,
hodaei2014parity,
poli2015selective,
helbig2020generalized}.
In that sense, it is a fundamental but largely unexplored question to what extent non-Hermitian quantum mechanics exhibits unique quantum nature distinct from classical systems.

\begin{figure}[tbp]
\begin{center}
\includegraphics[width=\columnwidth]{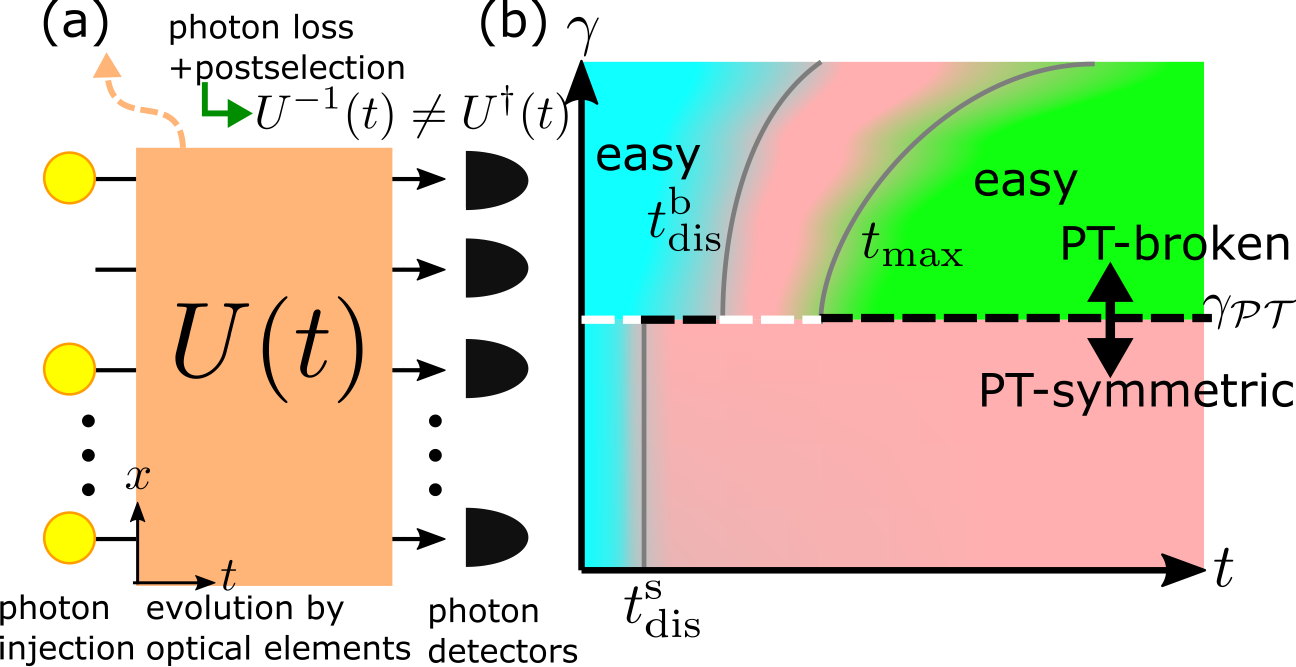}
\caption{(a) Schematic picture for the non-unitary boson sampling. 
Photons experience non-unitary dynamics by $U(t)\:(\neq[U^\dagger(t)]^{-1})$ through linear optical elements with loss effects, where we postselect cases for which all photons remain in the system. 
(b) Schematic phase diagram for sampling photon distributions with the algorithm that approximates photons to distinguishable particles. 
In blue and green regions, classical computers can efficiently sample the distribution of photons through the algorithm, where the latter is unique to non-unitary dynamics. 
Times $t_\mathrm{dis}^\mathrm{s}$ and $t_\mathrm{dis}^\mathrm{b}$ with $t_\mathrm{dis}^\mathrm{s}\ll t_\mathrm{dis}^\mathrm{b}$ are thresholds for the short-time dynamical distinguishability transition in the $\mathcal{PT}$-symmetric and $\mathcal{PT}$-broken phases, respectively. 
Time $t_\mathrm{max}$ is a threshold time for the long-time transition in the $\mathcal{PT}$-broken phase. 
 } 
\label{fig:boson-sampling_phase-diagram}
\end{center}
\end{figure}

In unitary quantum dynamics, unique quantum nature can be discussed by its computational complexity, which is investigated in the boson sampling problem \cite{aaronson2011computational,
aaronson2013the,
broome2013photonic,
crespi2013integrated,
spring2013boson,
tillmann2013experimental,
lund2014boson,
ticky2014stringent,
seshadreesan2015boson,
rahimi2015what,
aaronson2016bosonsampling,
latmiral2016towards,
rahimi2016sufficient,
hamilton2017gaussian,
he2017time,
loredo2017boson,
neville2017classical,
wang2017high,
deshpande2018dynamical,
oszmaniec2018classical,
garcia2019simulating,
huang2019simulating,
muraleedharan2019quantum,
roga2020classical,
zhong2020quantum,
oh2021classical,
chen2021non,
oh2022classical}. 
In this problem, probability distributions of photons in optical networks can be hard for classical computers to sample, which suggests quantum supremacy \cite{aaronson2013the}. 
Recent studies indicate that applicability and efficiency of classical algorithms for sampling photon distributions can be used to diagnose certain phases and states of photons obeying unitary evolution~\cite{seshadreesan2015boson,deshpande2018dynamical,muraleedharan2019quantum,oh2022classical}.

In this paper, we report novel transitions concerning the applicability of a classical algorithm which approximates photons to distinguishable particles in non-unitary boson sampling dynamics. 
We find that $\mathcal{PT}$-symmetry breaking, a unique phenomenon due to non-unitarity \cite{bender1998real,
bender2002complex,
bender2002generalized,
bender2003must,
mostafazadeh2003exact,
bender2007making,
guo2009observation,
ruter2010observation,
lin2011unidirectional,
chtchelkatchev2012stimulation,
miri2012large,
regensburger2012parity,
peng2014parity,
feng2014single,
hodaei2014parity,
mochizuki2016explicit,
ashida2017parity,
xiao2017observation,
longhi2018parity,
el2018non,
li2019observation,
longhi2019non,
kawasaki2020bulk,
mochizuki2020bulk,
takasu2020pt}, enhances classical regimes where the algorithm based on distinguishable particles can efficiently sample the distribution of photons.  
In a $\mathcal{PT}$-symmetric phase, there is a single dynamical transition as in isolated systems \cite{deshpande2018dynamical}, where the actual distribution ceases to be approximated by that of distinguishable particles. 
Upon $\mathcal{PT}$-symmetry breaking, the corresponding transition time is suddenly prolonged since dynamics of a photon is altered from a ballistic to a diffusive one. 
Remarkably, we discover an additional dynamical transition in the $\mathcal{PT}$-broken phase where the photon distribution is essentially determined by the dominant eigenmode of the non-unitary time-evolution operator, which makes photons distinguishable again. 
We also show that this long-time transition is generic for postselected non-unitary dynamics. 
In addition, we discover that the classical regime in the $\mathcal{PT}$-broken phase can also be captured by an algorithm based on the rank of matrices proposed in Ref. \cite{barvinok1996two}; the classical algorithm in Ref. \cite{barvinok1996two} efficiently obtains the distribution of many photons in the long-time regime of the $\mathcal{PT}$-broken phase, while this is not the case in the $\mathcal{PT}$-symmetric phase. 
Our model and its dynamical phase diagram are summarized in Fig. \ref{fig:boson-sampling_phase-diagram}.

Our transitions are characterized by the applicability of the algorithm based on distinguishable particles and the efficiency of the algorithm based on the rank of matrices. 
Such an idea of using a specific classical algorithm to characterize quantum phases has naturally been employed in the study of quantum matters. 
For example, various quantum phase transitions associated with entanglement are characterized by the efficiency of an algorithm based on matrix product states, such as the ground-state phase transition \cite{eisert2010colloquium}, many-body localization \cite{Friesdorf2015many,abanin2019colloquium}, and measurement-induced transition \cite{skinner2019measurement}. 
Instead of the matrix product states, we show that our distinct algorithm can capture the fundamental quantum/classical aspect of non-unitary dynamics and the $\mathcal{PT}$ symmetry breaking transition, since the difference between photons and distinguishable particles originates from quantum interference.

The rest of this paper is organized as follows. 
In Sec. \ref{sec:model_PT-breaking}, we introduce our model and explain that the model exhibits $\mathcal{PT}$ symmetry breaking transition. 
In Sec. \ref{sec:boson-sampling}, we give the setting of sampling bosons under non-unitary time evolution. 
In Sec. \ref{sec:transition_short-time}, we clarify that $\mathcal{PT}$ symmetry breaking prolongs transition times at which the probability distribution of photons deviates from that of distinguishable particles. This extends the region where the corresponding classical algorithm can sample the photon distribution efficiently. 
In Sec. \ref{sec:transition_long-time}, we show that $\mathcal{PT}$ symmetry breaking also leads to an emergence of a region where the classical algorithm based on distinguishable particles can sample photon distributions efficiently in the long-time regime. 
In Sec. \ref{sec:rank-reduction}, we reveal that another classical algorithm can obtain the photon distribution efficiently in the long-time regime of the $\mathcal{PT}$-broken phase. Put differently, this algorithm also captures the enhancement of classical natures due to $\mathcal{PT}$ symmetry breaking. 
Sections \ref{sec:discussion} and \ref{sec:conclusion} are devoted to discussion and conclusion, respectively.

\section{Model and $\mathcal{PT}$ symmetry breaking}
\label{sec:model_PT-breaking}
To consider free-boson dynamics with $\mathcal{PT}$-symmetry, we first introduce a model of discrete-time quantum walks, which are versatile platforms for exploring quantum statistics \cite{omar2006quantum,
pathak2007quantum,
peruzzo2010quantum,
mayer2011couting,
schreiber20122d,
cardano2015quantum,
wang2016repelling,
esposito2022quantum} and non-unitary dynamics \cite{mochizuki2016explicit,
xiao2017observation,
longhi2019non,
wang2019observation,
kawasaki2020bulk,
mochizuki2020bulk,
xiao2020non}. 
There, photons with two internal polarization states propagate through an optical network with $L$ sites, as detailed in Appendix \ref{sec:optical-elements}. 
In our one-dimensional system, the single-photon evolution operator for each timestep, parametrized by the rotation angles $\theta_1,\theta_2$ and gain/loss strength $\gamma$, reads
\begin{align}
    V=C(\theta_1/2)SG(\gamma)C(\theta_2)G(-\gamma)SC(\theta_1/2),
    \label{eq:floquet-operator_QW}
\end{align}
where 
\begin{align}
    &C(\theta)=I_x \otimes e^{-i\theta\sigma_2},\,
    G(\gamma)=I_x\otimes e^{\gamma\sigma_3},\nonumber\\
    S=&\sum_{x=1}^L\left(\begin{array}{cc}
    \ket{x+1}\bra{x} & 0 \\
    \hspace{-8mm}0 & \hspace{-8mm}\ket{x-1}\bra{x}
    \end{array}\right)=
    \sum_k\ket{k}\bra{k} \otimes e^{ik\sigma_3}
    \label{eq:operators}.
\end{align}
Here, the basis set is represented as $\hat{b}^\dagger_{x,\sigma}\ket{0}=\ket{x}\otimes\ket{\sigma}$ with horizontal/vertical polarization states written as $\ket{\sigma}=\ket{h}/\ket{v}$ being eigenstates of $\sigma_3$ with eigenvalues $+1$ and $-1$ respectively, where $x\in[1,2,\cdots,L]$ is the position with the periodic boundary condition, $\hat{b}_{x,\sigma}^\dagger$ is a creation operator of a photon with $x,\sigma$, and $\ket{0}$ is the vacuum. 
In addition, $\sigma_{1,2,3}$ are the Pauli matrices acting on the polarization, $I_x=\sum_{x=1}^L\ket{x}\bra{x}$, and the Fourier transform has been made for $S$ with a momentum basis $\ket{k}$. 
Single-photon dynamics for one timestep satisfies
\begin{align}
    \ket{\psi(t+1)}=V\ket{\psi(t)}/\sqrt{\bra{\psi(t)}V^\dagger V\ket{\psi(t)}}.
    \label{eq:dynamics_single-photon}
\end{align}
Non-unitary dynamics described by Eq. (\ref{eq:dynamics_single-photon}) has been investigated in photonic systems composed of optical elements, such as wave plates and beam splitters \cite{xiao2017observation,xiao2018higher,xiao2020non}. 
Such quantum optical systems are versatile platforms not only for investigating dynamics in open quantum systems \cite{xiao2017observation,xiao2018higher,xiao2020non} but also for exploring the boson sampling problem \cite{broome2013photonic,
crespi2013integrated,
tillmann2013experimental,
spring2013boson,
wang2017high}, which relates these two research fields.

Note that the polarization-dependent loss described by $G(\pm\gamma)$, which is different from usual boson sampling settings, is chosen such that the time-evolution operator respects $\mathcal{PT}$ symmetry~\cite{mochizuki2016explicit,xiao2017observation}. 
It is a fundamental research interest, especially in condensed matter physics, what type of symmetries leads to intriguing phase transitions. 
As we explain in the following, $\mathcal{PT}$ symmetric open quantum systems can exhibit a transition absent in isolated quantum systems, i.e., $\mathcal{PT}$ symmetry breaking. 
This indicates that $\mathcal{PT}$ symmetry breaking captures a qualitative distinction between isolated and open systems.

We define $\mathcal{PT}$ symmetry for a time-evolution operator $V$ as
\begin{align}
    (\mathcal{PT})V(\mathcal{PT})^{-1}=V^{-1},
    \label{eq:PT-symmetry_V}
\end{align}
where the symmetry operator $\mathcal{PT}$ includes the complex conjugation $\mathcal{K}$. 
When $V$ respects $\mathcal{PT}$ symmetry, there are two cases regarding its eigenstates,
\begin{align}
    V\ket{\phi_l^\mathrm{R}}=\lambda_l\ket{\phi_l^\mathrm{R}},\ \bra{\phi_l^\mathrm{L}}V=\lambda_l\bra{\phi_l^\mathrm{L}},
    \label{eq:eigen-equation}
\end{align} 
where $\lambda_l$ with $l=1,2,\cdots,M$ is an eigenvalue of $V$ and the eigenstates are bi-orthogonal as $\left\langle\phi_l^\mathrm{L}|\phi_m^\mathrm{R}\right\rangle=\delta_{lm}$. 
When an eigenstate $\ket{\phi_l^\mathrm{L/R}}$ satisfies $\mathcal{PT}$ symmetry, we have $|\lambda_l|=1$, while $\mathcal{PT}$ symmetry breaking of $\ket{\phi_l^\mathrm{L/R}}$ results in $|\lambda_l|\neq1$, that is,
\begin{align}
    \begin{cases}
    \mathcal{PT}\ket{\phi_l^\mathrm{L/R}}=\ket{\phi_l^\mathrm{L/R}}
    \rightarrow|\lambda_l|=1,\\
    \mathcal{PT}\ket{\phi_l^\mathrm{L/R}}\neq\ket{\phi_l^\mathrm{L/R}}
    \rightarrow|\lambda_l|\neq1.    
    \end{cases}
    \label{eq:PT-symmetry_state}
\end{align}
Here, the equality for the eigenstate is satisfied up to a phase factor. 
These two different behaviors define the $\mathcal{PT}$-symmetric and $\mathcal{PT}$-broken phases.
In the $\mathcal{PT}$-symmetric phase, all eigenstates respect $\mathcal{PT}$ symmetry, and all eigenvalues satisfy $|\lambda_l|=1$.
In the $\mathcal{PT}$-broken phase, some of the eigenstates break $\mathcal{PT}$ symmetry, and $|\lambda_l|\neq1$ for the corresponding eigenvalues.  
While we have explained $\mathcal{PT}$ symmetry for time-evolution operators and their eigenstates, this can be translated to Hamiltonians as $(\mathcal{PT})H(\mathcal{PT})^{-1}=H$ through $V=e^{-iH}$. 
In this case, $|\lambda_l|=1$ in the $\mathcal{PT}$-symmetric phase and $|\lambda_l|\neq1$ in the $\mathcal{PT}$-broken phase correspond to $\varepsilon_l\in\mathbb{R}$ and $\varepsilon_l\in\mathbb{C}$, respectively, where quasi-energies $\{\varepsilon_l=i\log(\lambda_l)\}$ are eigenvalues of the non-Hermitian Hamiltonian $H=i\log(V)$ \cite{mochizuki2016explicit}. 
In many cases, $\mathcal{PT}$-symmetric systems experience transitions from the $\mathcal{PT}$-symmetric phase to the $\mathcal{PT}$-broken phase as a parameter is altered. 

\begin{figure}[tbp]
\begin{center}
\includegraphics[width=\columnwidth]{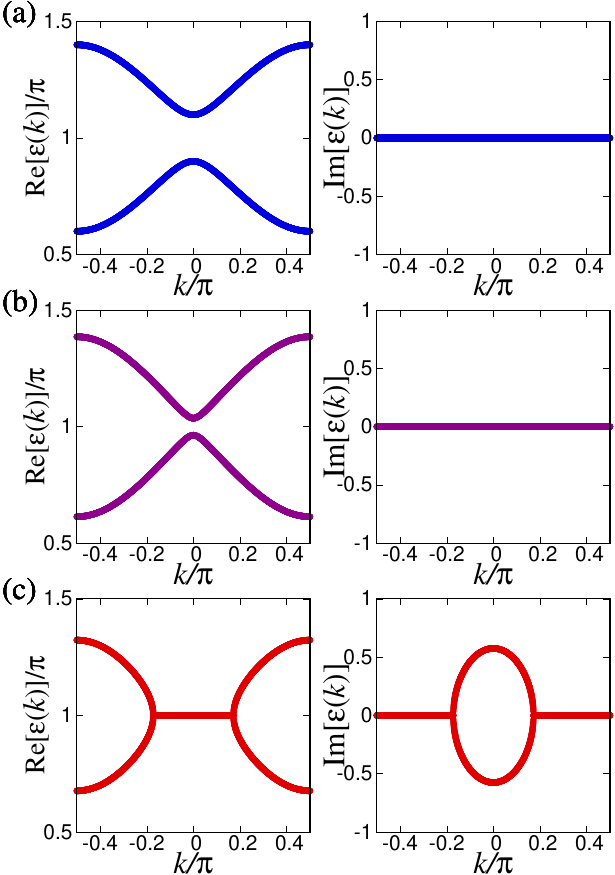}
\caption{ Quasi-energy dispersion relations of $\tilde{H}(k)=i\log[\tilde{V}(k)]$ with $\theta_1=0.65\pi,\,\theta_2=0.25\pi$, (a) $e^\gamma=1$ where $\tilde{V}(k)$ is unitary, (b) $e^\gamma=1.2$ in the $\mathcal{PT}$-symmetric phase, and (c) $e^\gamma=1.5$ in the $\mathcal{PT}$-broken phase. 
Some quasi-energies have non-zero imaginary parts in the $\mathcal{PT}$-broken phase, while all quasi-energies are real in the $\mathcal{PT}$-symmetric phase. }
\label{fig:dispersion}
\end{center}
\end{figure}

In systems with translational invariance, we consider $\mathcal{PT}$ symmetry and eigen-equations in momentum space through the Fourier transformation with $V=\sum_k\ket{k}\bra{k}\otimes\tilde{V}(k)$. 
When a $\mathcal{PT}$ symmetry operator is written as $\mathcal{PT}=\sum_x\ket{-x}\bra{x}\otimes\Omega$ with $\Omega$ acting on internal states, Eq. (\ref{eq:PT-symmetry_V}) corresponds to 
\begin{align}
    \Omega \tilde{V}(k) \Omega^{-1}=\tilde{V}^{-1}(k),
    \label{eq:PT-symmetry_V(k)}
\end{align}
which results in $\Omega\tilde{H}(k)\Omega^{-1}=\tilde{H}(k)$ with $\tilde{H}(k)=i\log[\tilde{V}(k)]$. 
In the present model $\tilde{V}(k)$ becomes 
\begin{align}
    \tilde{V}(k)=e^{-i\frac{\theta_1}{2}\sigma_2}
    e^{(ik+\gamma)\sigma_3}
    e^{-i\theta_2\sigma_2}
    e^{(ik-\gamma)\sigma_3}
    e^{-i\frac{\theta_1}{2}\sigma_2},
    \label{eq:V(k)}
\end{align}
and we can easily confirm Eq. (\ref{eq:PT-symmetry_V(k)}) with $\Omega=\sigma_3\mathcal{K}$ through $\sigma_3(e^{-i\theta\sigma_2})^\ast\sigma_3=e^{+i\theta\sigma_2}$, $\sigma_3(e^{+ik\sigma_3})^\ast\sigma_3=e^{-ik\sigma_3}$, and $\sigma_3(e^{\gamma\sigma_3})^\ast\sigma_3=e^{\gamma\sigma_3}$ \cite{mochizuki2016explicit}.
The eigenvalues are labeled by momentum $k$ and band indices $s=\pm$, where eigen-equations become 
\begin{align}
    \tilde{V}(k)\ket{\phi_\pm^\mathrm{R}(k)}
    =\lambda_\pm(k)\ket{\phi_\pm^\mathrm{R}(k)}
    \label{eq:eigen-equation_right},\\
    \bra{\phi_\pm^\mathrm{L}(k)}\tilde{V}(k)=
    \bra{\phi_\pm^\mathrm{L}(k)}\lambda_\pm(k).
    \label{eq:eigen-equation_left}
\end{align}
The relation between $\mathcal{PT}$ symmetry of eigenstates and the absolute values of eigenvalues is the same as Eq. (\ref{eq:PT-symmetry_state}), where $\mathcal{PT}$ is replaced with $\Omega$ and the label becomes $l=(k,s)$. 
The eigenvalues of $\tilde{V}(k)$ become
\begin{align}
    \lambda_\pm(k)=e^{-i\varepsilon_\pm(k)}=d(k)\pm\sqrt{d^2(k)-1},
    \label{eq:eigenvalue}
\end{align}
where $d(k)$ is defined as $d(k)=\cos(\theta_1)\cos(\theta_2)\cos(2k)
-\sin(\theta_1)\sin(\theta_2)\cosh(2\gamma)$.

Figure \ref{fig:dispersion} shows the real and imaginary parts of the quasi-energies with $\theta_1=0.65\pi$ and $\theta_2=0.25\pi$. 
Throughout this paper, we fix $\theta_1$ and $\theta_2$ at these values. 
Increasing the gain-loss parameter $\gamma$ with $\theta_1,\theta_2$ being fixed, the present model exhibits the transition from the $\mathcal{PT}$-symmetric phase to the $\mathcal{PT}$-broken phase. 
When the dissipation is weak as in Fig.~\ref{fig:dispersion} (b), all eigenstates preserve $\mathcal{PT}$ symmetry $\sigma_3\mathcal{K}\ket{\phi_\pm^\mathrm{L/R}(k)}=\ket{\phi_\pm^\mathrm{L/R}(k)}$ and thus $\varepsilon_\pm(k)$ have no imaginary part for arbitrary $k$, similar to the unitary case in (a). 
When the value of $\gamma$ is large, some of the eigenstates break $\mathcal{PT}$ symmetry $\sigma_3\mathcal{K}\ket{\phi_\pm^\mathrm{L/R}(k)}\neq\ket{\phi_\pm^\mathrm{L/R}(k)}$, resulting in non-zero imaginary parts of $\varepsilon_\pm(k)$ for some $k$, which is shown in Fig. \ref{fig:dispersion} (c). 
Since $\cos(\theta_1)\cos(\theta_2)<0$ and $\sin(\theta_1)\sin(\theta_2)>0$, the threshold value $\gamma_\mathcal{PT}$ for $\mathcal{PT}$ symmetry breaking is determined through $\lambda_-(k=0)=-1$, and we can obtain $\gamma_\mathcal{PT}=\cosh^{-1}\left[\{1+\cos(\theta_1)\cos(\theta_2)\}/\sin(\theta_1)\sin(\theta_2)\right]/2$, which leads to $e^{\gamma_\mathcal{PT}}\simeq1.22$. 
We note that the present model satisfies particle-hole symmetry $\tilde{V}^\ast(k)=\tilde{V}(-k)$ and thus $\varepsilon_\pm(k)=-\varepsilon_\pm^\ast(k)$, while this symmetry is not crucial to the transitions.

\section{Non-unitary boson sampling}
\label{sec:boson-sampling}
In the boson sampling dynamics, input $M$-mode bosons experience a linear transformation by an $M$-by-$M$ matrix $U(t)$, as schematically described in Fig. \ref{fig:boson-sampling_phase-diagram} (a). 
While previous works mainly focus on unitary $U(t)$, we here consider the non-unitary $\mathcal{PT}$-symmetric Floquet model introduced above and set \begin{align}
    U^\mathrm{T}(t)=V^t,
    \label{eq:time-evolution}
\end{align}
where $\mathrm{T}$ is the transpose. 
The basis of $U(t)$ is labeled by the positions and polarizations of a photon with an index $j=(x,\sigma)$ and thus $M=2L$. 
We can characterize the input state in two ways.
One way is to label the state by the index $j=(x,\sigma)$ for each photon: the state is given by $\{\mathsf{In}\}=(\mathsf{In}_{1},\cdots,\mathsf{In}_n)$, where $\mathsf{In}_p=(x_p^\mathrm{in},\sigma_p^\mathrm{in})$ is the corresponding index for the $p$th input photon, and $n$ is the total number of photons. 
The other way is to characterize the input state by the number of photons at each $j$, i.e., $\{n^\mathrm{in}\}=(n^\mathrm{in}_1,\cdots,n^\mathrm{in}_M)$ with $\sum_jn^\mathrm{in}_j=n$. 
As shown in Appendix \ref{sec:optical-elements}, a normalized state $\prod_{p=1}^n\hat{b}_{\mathsf{In}_p}^\dagger\ket{0}/\prod_{j=1}^M\sqrt{n_j^\mathrm{in}!}$ is transformed as
\begin{align}
    \frac{\prod_{p=1}^n
    \hat{b}_{\mathsf{In}_p}^\dagger\ket{0}}
    {\prod_{j=1}^M\sqrt{n_j^\mathrm{in}!}}
    \rightarrow
    \frac{\prod_{p=1}^n
    \left[\sum_{j=1}^MU_{\mathsf{In}_pj}(t)
    \hat{b}_j^\dagger\right]\ket{0}}
        {\sqrt{N(t)}\prod_{j=1}^M\sqrt{n_j^\mathrm{in}!}},
    \label{eq:creation-operator}
\end{align}
where $N(t)$ is a normalization factor, which is unity only when $U(t)$ is unitary. 
For our case, $U(t)$ is non-unitary unless $e^\gamma=1$.

From Eq. (\ref{eq:creation-operator}), the output probability distribution of photons, characterized by $\{n^\mathrm{out}\}=(n^\mathrm{out}_1,\cdots,n^\mathrm{out}_M)$ with $\sum_jn^\mathrm{out}_j=n$, becomes
\begin{align}
    P(\{n^\mathrm{in}\},\{n^\mathrm{out}\},t)
    =\frac{|\mathrm{Per}[W(t)]|^2}
    {N(t)\prod_{j=1}^Mn^\mathrm{in}_j!n^\mathrm{out}_j!}.
    \label{eq:probability-distribution}
\end{align}
Here, $\mathrm{Per}[W(t)]$ is the permanent of an $n$-by-$n$ matrix $W(t)$ defined as $W_{pq}(t)=U_{\mathsf{In}_p\mathsf{Out}_q}(t)$, where $\mathsf{Out}_q=(x_q^\mathrm{out},\sigma_q^\mathrm{out})$ is the label for the $q$th output photon.
Note that Eq.~\eqref{eq:probability-distribution} has the same form with the unitary case~\cite{scheel2004permanents} except for the nomalization factor $N(t)$. The essence of the boson sampling problem is that   $P(\{n^\mathrm{in}\},\{n^\mathrm{out}\},t)$ can be  hard to sample by a classical computer. For example, Ref.~\cite{aaronson2013the} provided a plausible conjecture based on a polynomial hierarchy that the problem is \#P-hard when $U(t)$ is a Gaussian random unitary matrix. 
On the other hand, constraints on $U(t)$, such as the positivity of matrix elements or the generatability by short-time dynamics of local Hamiltonians, result in distributions that are efficiently sampled \cite{aaronson2011computational,deshpande2018dynamical}. 

Employing an algorithm that approximates photons to distinguishable particles, we characterize dynamical behavior and $\mathcal{PT}$ symmetry breaking of the open quantum system. 
To this end, we evaluate the $L_1$-distance between $P(\{n^\mathrm{in}\},\{n^\mathrm{out}\},t)$ and the reference distribution $P_\mathrm{dis}(\{n^\mathrm{in}\},\{n^\mathrm{out}\},t)$,
\begin{align}
    \eta(t)=\sum_{\{n^\mathrm{out}\}}
    \left|P(\{n^\mathrm{out}\},t)-P_\mathrm{dis}(\{n^\mathrm{out}\},t)\right|,
    \label{eq:difference}
\end{align}
where $\{n^\mathrm{in}\}$ is omitted for brevity throughout the manuscript, if there is no confusion.  
Here, $P_\mathrm{dis}(\{n^\mathrm{out}\},t)$ is the probability distribution of distinguishable particles
\begin{align}
    P_\mathrm{dis}(\{n^\mathrm{in}\},\{n^\mathrm{out}\},t)
    =\frac{\sum_\omega
    \prod_{p=1}^n|U_{\mathsf{In}_p\mathsf{Out}_{\omega(p)}}(t)|^2}
    {N_\mathrm{dis}(t)\prod_{j=1}^Mn^\mathrm{in}_j!n^\mathrm{out}_j!},
    \label{eq:probability-distribution_distinguishable}
\end{align} 
where $\omega$ in the sum is taken over all permutations for the output state, and $N_\mathrm{dis}(t)=\prod_{p=1}^n\bra{\mathsf{In}_p}U^\ast(t)U^\mathrm{T}(t)\ket{\mathsf{In}_p}$ is the normalization factor. 
Since sampling $P_\mathrm{dis}(\{n^\mathrm{out}\},t)$ can be carried out within the polynomial computational time of $n$ \cite{deshpande2018dynamical}, if $\eta(t)<\delta$ is satisfied with a constant $\delta$, sampling $P(\{n^\mathrm{out}\},t)$ is defined to be classically easy within the precision $\delta$. 
In the following, we use $\eta(t)$ as an order parameter.  

We notice that our criterion $\eta(t)<\delta$ for ``efficient boson sampling" by classical computers is different from that in the original boson sampling paper \cite{aaronson2011computational}. 
Reference \cite{aaronson2011computational} considers classical algorithms that sample a probability distribution $P'(\{n^\mathrm{out}\},t)$ with a polynomial computational time of $n$ and $1/\delta'$, when we require that the $L_1$ distance between $P(\{n^\mathrm{out}\},t)$ and $P'(\{n^\mathrm{out}\},t)$ is smaller than $\delta'$. 
Then, the boson sampling problem is efficiently carried out by classical computers if such an algorithm exists. 
In contrast, in our definition, the threshold $\delta$ is fixed and not controllable; we cannot make $\delta$ smaller by increasing the computational time since the computational time of the algorithm based on distinguishable particles is independent of $\delta$. 
Nevertheless, our criterion is sufficient for discussing quantum and classical natures of the non-unitary dynamics of photons, where the deviation of photon distributions from that of distinguishable particles originates from quantum interference of photons.

\section{Distinguishability transition in the short-time regime}
\label{sec:transition_short-time}
Let us first discuss the short-time dynamical transition, which occurs when we place the input photons well separated from one another with the distance $L/n$. 
As we explain below, by choosing such an initial state, $\mathcal{PT}$ symmetry breaking can be diagnosed through distinguishability of photons in early times. 
Actually, these photons are approximately regarded as distinguishable particles for small $t$, where quantum interference is negligible. 
Then, $P(\{n^\mathrm{in}\},\{n^\mathrm{out}\},t)$ is approximated by  $P_\mathrm{dis}(\{n^\mathrm{in}\},\{n^\mathrm{out}\},t)$, which can be sampled efficiently by classical computers through sampling single-photon distributions \cite{deshpande2018dynamical}.

Due to time evolution, photons start to interfere and are no longer approximated by distinguishable particles.
Then, a dynamical distinguishability transition occurs; as time evolves, the system enters from a phase where $P(\{n^\mathrm{out}\},t)$ is classically easy to sample approximately through $P_\mathrm{dis}(\{n^\mathrm{out}\},t)$ into a phase where the classical algorithm using $P_\mathrm{dis}(\{n^\mathrm{out}\},t)$ is not applicable. 
The former and latter phases correspond to the blue and red regions in Fig. \ref{fig:boson-sampling_phase-diagram} (b), respectively. 
The order parameter for the transition is given by $\eta(t)$.

\begin{figure}[tbp]
\begin{center}
\includegraphics[width=\columnwidth]{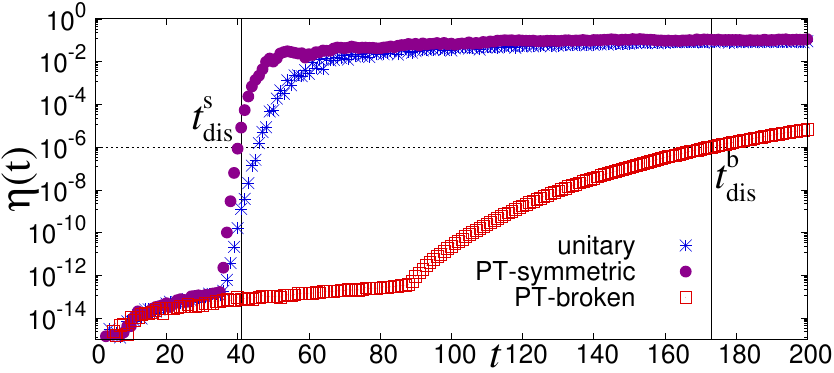}
\caption{$L_1$-distance between the actual probability distribution of photons $P(\{n^\mathrm{in}\},\{n^\mathrm{out}\},t)$ in Eq.  (\ref{eq:probability-distribution}) and that for distinguishable particles $P_\mathrm{dis}(\{n^\mathrm{in}\},\{n^\mathrm{out}\},t)$ in Eq.~(\ref{eq:probability-distribution_distinguishable}).
After the threshold time $t_\mathrm{dis}^\mathrm{s/b}$, the two distributions differ as $\eta(t)>\delta\:(=10^{-6})$, which defines the short-time distinguishability transition.
We find that $t_\mathrm{dis}^\mathrm{b}$ for the $\mathcal{PT}$-broken phase is much larger than $t_\mathrm{dis}^\mathrm{s}$ for the $\mathcal{PT}$-symmetric phase. 
The blue asterisks, purple circles, and red squares respectively correspond to $e^\gamma=1,\,1.2$, and $1.5$. 
The lowest value around $10^{-14}$ is a numerical artifact. 
The number of photons is $n=3$, and the initial state is $\hat{b}^\dagger_{L/6,h}\hat{b}^\dagger_{L/2,v}\hat{b}^\dagger_{5L/6,v}\ket{0}$ with $L=300$. 
The rotation angles are $\theta_1=0.65\pi$ and $\theta_2=0.25\pi$, with which the threshold gain-loss parameter $\gamma_\mathcal{PT}$ for $\mathcal{PT}$ symmetry breaking becomes $e^{\gamma_\mathcal{PT}}\simeq1.22$.}
\label{fig:difference_short}
\end{center}
\end{figure}

Figure \ref{fig:difference_short} shows numerically obtained $L_1$-norm $\eta(t)$ with $n=3$ and the initial distance between photons being $L/n=100$. 
We actually find the transition explained above for both $\mathcal{PT}$-symmetric and $\mathcal{PT}$-broken phases.
The threshold times denoted by $t_\mathrm{dis}^\mathrm{s}$ and $t_\mathrm{dis}^\mathrm{b}$ for these phases are defined as  $t_\mathrm{dis}^\mathrm{s/b}=\mathrm{min}\: t\text{\quad s.t.\quad}\eta(t)>\delta$  for small $\delta>0$, which describes the precision of the approximation.

Surprisingly, from Fig. \ref{fig:difference_short}, we discover that the threshold time $t_\mathrm{dis}^\mathrm{b}$  is much larger than $t_\mathrm{dis}^\mathrm{s}$. 
To understand the prolongation of the threshold time upon $\mathcal{PT}$ breaking, let us evaluate the time scale for the initially separated photons to start to interfere using single-photon dynamics of non-unitary quantum walks. 
As shown in Appendix \ref{sec:ballistic-diffusive}, we find that the quantum walk of a photon exhibits the ballistic behavior in the $\mathcal{PT}$-symmetric phase, which means that the standard deviation of $x$ behaves as $\sqrt{\left<x^2(t)\right>-\left<x(t)\right>^2} \propto t$, where $\left<x^m(t)\right>=\sum_{x,\sigma}x^m|\psi(x,\sigma,t)|^2$ with $\ket{\psi(t)}=\sum_{x,\sigma}\psi(x,\sigma,t)\ket{x}\otimes\ket{\sigma}$. 
This ballistic dynamics leads to $t_\mathrm{dis}^\mathrm{s}\propto L/n$. 
This behavior is similar to that found in unitary dynamics~\cite{deshpande2018dynamical}. 
We note that the dependence of $t_\mathrm{dis}^\mathrm{s}$ on $\gamma$ is weak compared to the dependence of $t_\mathrm{dis}^\mathrm{b}$ on $\gamma$. 

On the other hand, in the $\mathcal{PT}$-broken phase, we can show that the single-photon probability distribution becomes asymptotically Gaussian, and the standard deviation of $x$ behaves diffusive as
\begin{align}
    \sqrt{\left<x^2(t)\right>-\left<x(t)\right>^2}=\sqrt{\frac{Dt}{2}},
    \label{eq:deviation_PT-broken}
\end{align}
where the diffusion constant is $D=\left|\frac{d^2}{dk^2}\varepsilon_-(k)\right|_{k=0}=|4\cos(\theta_1)\cos(\theta_2)/\sqrt{d^2(k=0)-1}|$. 
In this case, $t_\mathrm{dis}^\mathrm{b}$ is proportional to $(L/n)^2$, which results in $t_\mathrm{dis}^\mathrm{b} \gg t_\mathrm{dis}^\mathrm{s}$ for large $L$. 
Indeed, as shown in Appendix \ref{sec:ballistic-diffusive}, we can evaluate $\eta(t)$ as
\begin{align}
    \eta(t)\leq4n\exp\left(-\frac{L^2}{2n^2Dt}\right),
    \label{eq:inequality_short-time}
\end{align}
in the short-time regime of the $\mathcal{PT}$-broken phase. 
Equation (\ref{eq:inequality_short-time}) leads to
\begin{align}
    t_\mathrm{dis}^\mathrm{b}
    >\frac{L^2}{2n^2D|\log(\delta/4n)|}.
    \label{eq:critical-t_short-time}
\end{align}
Since $1/D\propto e^{2\gamma}$ for large $\gamma$, $t_\mathrm{dis}^\mathrm{b}$ exhibits an exponential growth as a function of $\gamma$, as schematically described in Fig. \ref{fig:boson-sampling_phase-diagram} (b).

For the derivation of the above results, we only use the Fourier analysis and typical dispersion relations in $\mathcal{PT}$-symmetric systems. 
Therefore, the ballistic-diffusive transition due to $\mathcal{PT}$-symmetry breaking and the resulting prolongation of the threshold time for the dynamical distinguishability transition exist in a wide range of $\mathcal{PT}$-symmetric systems with translation invariance. 

We note that diffusive behavior and a short time transition of photons are also discussed in Ref. \cite{oh2022classical}, where dynamics in Haar-random unitary optical circuits is explored. 
However, in the case of our translational invariant system, diffusive dynamics does not originate from randomness but $\mathcal{PT}$ symmetry breaking. 
As another unique property in the present work, the $\mathcal{PT}$ symmetric non-unitary model exhibits both ballistic and diffusive behavior just by the change of the parameters, where the former and latter correspond to $\mathcal{PT}$-symmetric and $\mathcal{PT}$-broken phases, respectively.

\section{Distinguishability transition in the long-time regime}
\label{sec:transition_long-time}
Let us next focus on the long-time regime in the $\mathcal{PT}$-broken phase. 
We discover a novel dynamical transition absent in unitary dynamics, where the photons become distinguishable again and thus the corresponding classical algorithm can sample photon distributions efficiently. 
To this end, we carry out the spectral decomposition of $V^t$. 
In the $\mathcal{PT}$-broken phase, only one eigenstate becomes dominant in the long run, and thus the non-unitary matrix $U^\mathrm{T}(t)=V^t$ can be written as
\begin{align}
    U^\mathrm{T}(t)\simeq\lambda_\mathrm{max}^t\ket{\phi_\mathrm{max}^\mathrm{R}}\bra{\phi_\mathrm{max}^\mathrm{L}},
    \label{eq:maximal-state}
\end{align}
where $\ket{\phi_\mathrm{max}^\mathrm{R}}$ and $\ket{\phi_\mathrm{max}^\mathrm{L}}$ are respectively right and left eigenstates of $V$ for the largest eigenvalue $|\lambda_\mathrm{max}|\equiv\max_l|\lambda_l|=\mathrm{max}_{s,k}|\lambda_s(k)|=|\lambda_-(k=0)|$. 
On the basis of Eq. (\ref{eq:maximal-state}), the permanent of $W(t)$ becomes
\begin{align}
    \mathrm{Per}[W^*(t)] \simeq \lambda_\mathrm{max}^{nt} n!
    \prod_{p=1}^n\langle \mathsf{Out}_p
    |\phi_\mathrm{max}^\text{R} \rangle
    \langle\phi_\mathrm{max}^\text{L}
    |\mathsf{In}_p\rangle.
    \label{eq:permanent_dominant}
\end{align}
The normalization factor is obtained through Eq. (\ref{eq:probability-distribution}) and $\sum_{\{n^\mathrm{out}\}}P(\{n^\mathrm{in}\},\{n^\mathrm{out}\},t)=1$, and thus the probability distribution for large $t$ approaches 
\begin{align}
    P_\mathrm{max}(\{n^\mathrm{out}\})=\frac{\prod_{p=1}^n|\langle\mathsf{Out}_p|\phi_\mathrm{max}^\text{R}\rangle|^2}{N_\mathrm{max}\prod_{j=1}^Mn^\mathrm{out}_j!},
    \label{eq:P_max}
\end{align}
where $N_\mathrm{max}=\left\langle\phi_\mathrm{max}^\mathrm{R}|\phi_\mathrm{max}^\mathrm{R}\right\rangle^n/n!$. 
We can show that $\lim_{t\rightarrow\infty}P_\mathrm{dis}(\{n^\mathrm{in}\},\{n^\mathrm{out}\},t)$ and $P_\mathrm{max}(\{n^\mathrm{out}\})$ become the same if $U^\mathrm{T}(t)$ is written as Eq. (\ref{eq:maximal-state}), which indicates that $P_\mathrm{max}(\{n^\mathrm{out}\})$ is easy to sample with the classical algorithm based on distinguishable particles.

After a sufficiently long time, a dynamical transition occurs to a phase where sampling $P(\{n^\mathrm{out}\},t)$ is classically easy approximately through sampling $P_\mathrm{dis}(\{n^\mathrm{out}\},t)$ from a phase where the approximation to $P_\mathrm{dis}(\{n^\mathrm{out}\},t)$ is impossible and the algorithm based on distinguishable particles cannot be applied. 
The two phases respectively correspond to the green and red regions in Fig. \ref{fig:boson-sampling_phase-diagram} (b).

\begin{figure}[tbp]
\begin{center}
\includegraphics[width=\columnwidth]{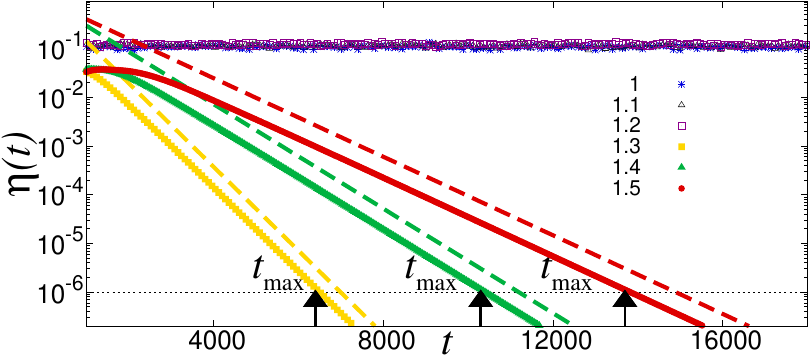}
\caption{$L_1$-distance between $P(\{n^\mathrm{out}\},t)$ and $P_\mathrm{dis}(\{n^\mathrm{out}\},t)$.  
After the threshold time $t \geq t_\mathrm{max}$ in the $\mathcal{PT}$-broken phase, $\eta(t)<\delta$ holds and thus the classical algorithm based on $P_\mathrm{dis}(\{n^\mathrm{out}\},t)$ can efficiently sample $P(\{n^\mathrm{out}\},t)$, which defines the long-time distinguishability transition. 
The dashed lines show $e^{-2\Delta t}$, which confirms Eq. (\ref{eq:critical-t_long-time}). 
The blue, black, purples, yellow, green, and red symbols correspond to $e^\gamma=1,\,1.1,\,1.2,\,1.3,\,1.4$, and $1.5$, respectively. 
The initial condition and parameters are the same as those in Fig. \ref{fig:difference_short}. 
The long-time distinguishability transition is absent in the $\mathcal{PT}$-symmetric phase for $e^\gamma=1.1,1.2$ and unitary dynamics without loss for $e^\gamma=1$. }
\label{fig:difference_long}
\end{center}
\end{figure}

Figure \ref{fig:difference_long} shows that $\eta(t)$ actually converges to zero for large $t$ in the $\mathcal{PT}$-broken phase after the threshold time defined as $t_\mathrm{max}=\min t \text{\quad s.t.\quad} \eta(t)<\delta$ under a constraint $t>t_\mathrm{dis}^\mathrm{b}$. 
The broken lines in Fig. \ref{fig:difference_long} show $e^{-2 \Delta t}$, where a gap $\Delta=D(2\pi/L)^2/2$ is defined as the difference between the largest value and the second largest value of $\mathrm{Im}[\varepsilon_-(k)]$.
From this, we find that the relaxation is governed by $\Delta$. 
Indeed, as detailed in Appendix \ref{sec:evaluation_tm}, we can show 
\begin{align}
    \eta(t)\leq32n^2e^{-2 \Delta t}
    \label{eq:inequality_long-time}
\end{align}
in the long-time regime of the $\mathcal{PT}$-broken phase. 
Equation (\ref{eq:inequality_long-time}) leads to
\begin{align}
    t_\mathrm{max}<
    \frac{|\log(\delta/32n^2)|}{2\Delta}=\frac{|\log(\delta/32n^2)|}{4D\pi^2}L^2,
    \label{eq:critical-t_long-time}
\end{align}
which means that sampling bosons is inevitably easy for $t>|\log(\delta/32n^2)|L^2/4D\pi^2$.
We argue that $\log(t_\mathrm{max})$ is proportional to $\gamma$, as schematically described in Fig. \ref{fig:boson-sampling_phase-diagram} (b), since $1/D\propto e^{2\gamma}$ if $\gamma$ is large.

The easiness of sampling the photon probability distribution for long times is not restricted to the present model and $\mathcal{PT}$-symmetric systems unless $|\lambda_\mathrm{max}|$ is degenerate. 
Therefore, our result indicates that the long-time dynamical transitions, at which the classical algorithm based on distinguishable particles becomes applicable to sample photon distributions, ubiquitously occur in postselected non-unitary quantum systems. 
We stress that such a transition is absent in unitary dynamics of isolated quantum systems. 

As mentioned above, we define dynamical transitions through the fixed algorithm, which approximates photons to distinguishable particles, while it is not proved if the problem is classically hard for arbitrary algorithms in the red region of Fig. \ref{fig:boson-sampling_phase-diagram} (b). 
Therefore, even if classically easy algorithms were found in the red region, the transitions characterized by $t_\mathrm{dis}^{\mathrm{s/b}}$ and $t_\mathrm{max}$ are well-defined from the closeness of $P(\{n^\mathrm{in}\},\{n^\mathrm{out}\},t)$ to $P_\mathrm{dis}(\{n^\mathrm{in}\},\{n^\mathrm{out}\},t)$.

\section{Dynamical transition through reduction of the matrix rank}
\label{sec:rank-reduction}
We can also define long-time dynamical transitions on the basis of the efficiency for a classical algorithm proposed in Ref. \cite{barvinok1996two}, which focuses on the matrix rank. 
While the algorithm \cite{barvinok1996two} we consider in this section is different from the one employed in Secs. \ref{sec:transition_short-time} and \ref{sec:transition_long-time}, the role of $\mathcal{PT}$ symmetry breaking is quite similar in both cases; the region where a classical algorithm is efficient emerges due to $\mathcal{PT}$ symmetry breaking. 

It is known that we can compute $\mathrm{Per}[W(t)]$ with a computational time $\mathcal{O}[f(n)]$ with $f(n)=\max\left[n^3,\frac{(n+r-1)!}{n!(r-1)!}\right]$ when we can approximately regard the rank of the $n \times n$ matrix $W(t)$ as $r\,(\leq n)$ \cite{barvinok1996two}. 
In the $\mathcal{PT}$-symmetric phase, $r=n$ results in the exponentially long computational time with respect to $n$. 
In the $\mathcal{PT}$-broken phase, some eigenstates corresponding to large $|\lambda_l|$ become dominant as time evolves, which results in $r \ll n$ and polynomial computational time $\mathcal{O}(n^{r-1})$ or $\mathcal{O}(n^3)$. 
We can thus find that the efficiency of the algorithm is again significantly altered by $\mathcal{PT}$ symmetry breaking.
Note that the phase diagram defined from the algorithm in Ref. \cite{barvinok1996two} is different from Fig. \ref{fig:boson-sampling_phase-diagram}(b) in that the blue region is absent and threshold times are different from $t_\mathrm{max}$. 

If we sort eigenvalues of $V$ as $|\lambda_1|\geq|\lambda_2|\geq\cdots\geq|\lambda_M|$ and neglect lower order terms with $l<r$, we can approximate $U(t)$ as
\begin{align}
    U^\mathrm{T}(t)\simeq\sum_{l=1}^r\lambda_l^t
    \ket{\phi_l^\mathrm{R}}\bra{\phi_l^\mathrm{L}}.
    \label{eq:U_low-rank}
\end{align}
In the right-hand side of Eq. (\ref{eq:U_low-rank}), the rank of each term in the sum is $1$, which indicates that the rank of $U(t)$ is approximately smaller than or equal to $r$. 
This is because $\mathrm{rank}(A+B)\leq\mathrm{rank}(A)+\mathrm{rank}(B)$ is satisfied for arbitrary matrices $A$ and $B$. 
Thanks to a similar discussion, the rank of $W(t)$ is also smaller than or equal to $r$ approximately. 

Here, we put a cutoff $\tilde{\delta}$ and neglect components that satisfy $|\lambda_l/\lambda_1|^t<\tilde{\delta}$. 
In the present model, the dispersion relation of the growing modes can be approximated as an imaginary quadratic function $\varepsilon_-(k)\simeq\varepsilon(k=0)-iDk^2/2$ near $k=0$, as detailed in Appendix \ref{sec:ballistic-diffusive}. 
Thus, we can evaluate the ratio as $|\lambda_l/\lambda_1|=|\lambda_-(k)/\lambda_-(k=0)|\simeq\exp\left(-Dk^2/2\right)$, where $l=kL/\pi$ for $k>0$ and $l=|kL/\pi|+1$ for $k<0$. 
When a momentum $\tilde{k}$ satisfies
\begin{align}
    \exp\left(-\frac{D\tilde{k}^2}{2}t\right)\simeq\tilde{\delta},
    \label{eq:cutoff_momentum}
\end{align}
the number of eigenstates that satisfy $|\lambda_-(k)/\lambda_-(k=0)|^t>\tilde{\delta}$ becomes $2|\tilde{k}|/(2\pi/L)$, which corresponds to $r$. 
If the rank of $W(t)$ can be approximated by an $n$-independent value $r_\mathrm{c}<n$, we can efficiently compute $\mathrm{Per}[W(t)]$ with the algorithm in Ref. \cite{barvinok1996two}. 
Here, we define a threshold time $t_\mathrm{rank}$ as the smallest time at which we can compute $P(\{n^\mathrm{in}\},\{n^\mathrm{out}\},t)$ by approximating the rank of $U(t)$ by $r_\mathrm{c}$ within the precision $\tilde{\delta}$. 
Then, we can evaluate $t_\mathrm{rank}$ as 
\begin{align}
    t_\mathrm{rank}\simeq\frac{2|\log(\tilde{\delta})|}{D\pi^2}
    \left(\frac{L}{r_\mathrm{c}}\right)^2 \gg\frac{2|\log(\tilde{\delta})|}{D\pi^2}
    \left(\frac{L}{n}\right)^2.
    \label{eq:t_rank}
\end{align}
While a more detailed evaluation of $t_\mathrm{rank}$ based on comparing $P(\{n^\mathrm{in}\},\{n^\mathrm{out}\},t)$ and the distribution obtained through the algorithm in \cite{barvinok1996two} may result in corrections to the right-hand side of Eq. (\ref{eq:t_rank}), such as $\log(n)$ in Eq. (\ref{eq:critical-t_long-time}), we believe that such effects are negligible for sufficiently small $\tilde{\delta}$ and large $L/n$.

\section{Discussions}
\label{sec:discussion}
While we have focused on \textit{sampling} probability distributions of photons, \textit{computing} $P_\mathrm{dis}(\{n^\mathrm{in}\},\{n^\mathrm{out}\},t)$ for small and large $t$ is also classically easy. 
The easiness for $P_\mathrm{dis}(\{n^\mathrm{out}\},t)$ with small $t$ is because the computational time becomes $\mathcal{O}(n)$ due to the significant simplification of the sum over $\omega$ in  Eq.~\eqref{eq:probability-distribution_distinguishable}; that is, the matrix elements of $U(t)$ for small $t$ almost vanish for the input and output states whose photon configurations are not spatially close. 
The easiness for computing $P_\mathrm{dis}(\{n^\mathrm{out}\},t)$ with large $t$, which approaches to $P_\mathrm{max}(\{n^\mathrm{out}\})$, is due to the fact that it only takes $\mathcal{O}(M^3)$ computational time by numerically diagonalizing $V$ to compute $P_\mathrm{max}(\{n^\mathrm{out}\})$. 

We also note that in blue and green regions of Fig. \ref{fig:boson-sampling_phase-diagram} (b), the postselection probability of bosons has little effect on the efficiency of sampling photon distributions based on classical computers. 
This is because, while the postselection probability alters normalization factors of quantum states, we can normalize distributions of distinguishable particles by computing $N_\mathrm{dis}(t)$ within a polynomial computational time of $n$ using single-particle distributions. 
Note that the postselection raises the cost of the actual experiment since the probability that all photons remain in the system decreases exponentially as time evolves.

We here remind that the meanings of ``efficient boson sampling" in Ref. \cite{aaronson2011computational} and the present work have some difference. 
As mentioned in the last paragraph of Sec. \ref{sec:boson-sampling}, our algorithm based on distinguishable particles does not have a controllable error parameter that is considered in the original boson sampling paper \cite{aaronson2011computational}.  
We regard the boson sampling as classically efficient when the classical algorithm which approximates photons to distinguishable particles can sample the distribution of photons efficiently within the precision $\delta$, where $\delta$ is fixed and not controllable. 
While the classical algorithm based on distinguishable particles does not have the controllable error parameter, the inequalities that we derived in the present work are related to the precision of the sampling. 
For instance, if we want to sample photon distributions using the classical algorithm at time $t$ in the long-time regime of the $\mathcal{PT}$-broken phase, Eq. (\ref{eq:inequality_long-time}) leads to a restriction on the precision; choosing the precision $\delta'$ such that $\delta'<32n^2e^{-2 \Delta t}$ may lead to incorrect sampling with $\eta(t)>\delta'$, which deviates from the required precision; in contrast, if we choose $\delta'$ larger than $32n^2e^{-2 \Delta t}$, such $\delta'$ guarantees $\eta(t)\leq\delta'$. 
In the short-time regime of the $\mathcal{PT}$-broken phase, Eq. (\ref{eq:inequality_short-time}) plays the same role. 

We briefly discuss distinctions between our work and previous works on the lossy or non-unitary boson sampling in Refs. \cite{oszmaniec2018classical,
garcia2019simulating,
huang2019simulating,
oh2021classical,
chen2021non}.
While Refs. \cite{oszmaniec2018classical,
garcia2019simulating,
huang2019simulating,
oh2021classical} considered the case where the number of photons decreases, our result shows that the classical algorithms we employed become efficient for sampling photon distributions in the $\mathcal{PT}$-broken phase even when the number of photons is conserved. 
Also, while the mechanism for the classically easy sampling is attributed to the localization of photons in Ref. \cite{chen2021non}, it is due to the closeness between the actual distribution of photons and that of distinguishable particles in our work. 
Furthermore, the main result of our work, i.e., the transition associated with the applicability and/or efficiency of classical algorithms due to the dominant eigenstate, was not discussed in Refs. \cite{oszmaniec2018classical,
garcia2019simulating,
huang2019simulating,
oh2021classical,
chen2021non}.

\section{Conclusion}
\label{sec:conclusion}
We have shown that $\mathcal{PT}$ symmetry breaking, a transition unique to open systems, enhances classical regimes where photons are distinguishable and makes two classical algorithms efficient to sample or compute the probability distribution of photons in non-unitary dynamics. 
Regarding the short-time transition at which the distribution of photons deviates from that for distinguishable particles, $\mathcal{PT}$ symmetry breaking prolongs the threshold time through the ballistic-diffusive transition of single-photon dynamics. 
Remarkably, $\mathcal{PT}$ symmetry breaking leads to the additional long-time dynamical transitions at which the classical algorithms become efficient, which is absent in isolated systems. 
This long-time transition appears in a wide range of non-unitary dynamics.

As future works, it should be interesting to explore efficiency and applicability of classical algorithms in other types of open systems, such as many-body interacting systems and Markovian open systems described by Gorini–Kossakowski–Sudarshan–Lindblad master equation. 
It will also be intriguing to study relations between transitions in open quantum systems and other indicators of the complexity, such as the entanglement entropy and quantum circuit complexity.

\begin{acknowledgements}
We thank Tomotaka Kuwahara, Hideaki Obuse, and Yutaka Shikano for fruitfull discussions. 
\end{acknowledgements}

\appendix
\section{Components of $V$ realized by optical elements and postselections}
\label{sec:optical-elements}
We here show that our model describes postselected non-unitary dynamics of photons passing through optical elements.

\begin{figure}[tbp]
\begin{center}
\includegraphics[width=\columnwidth]{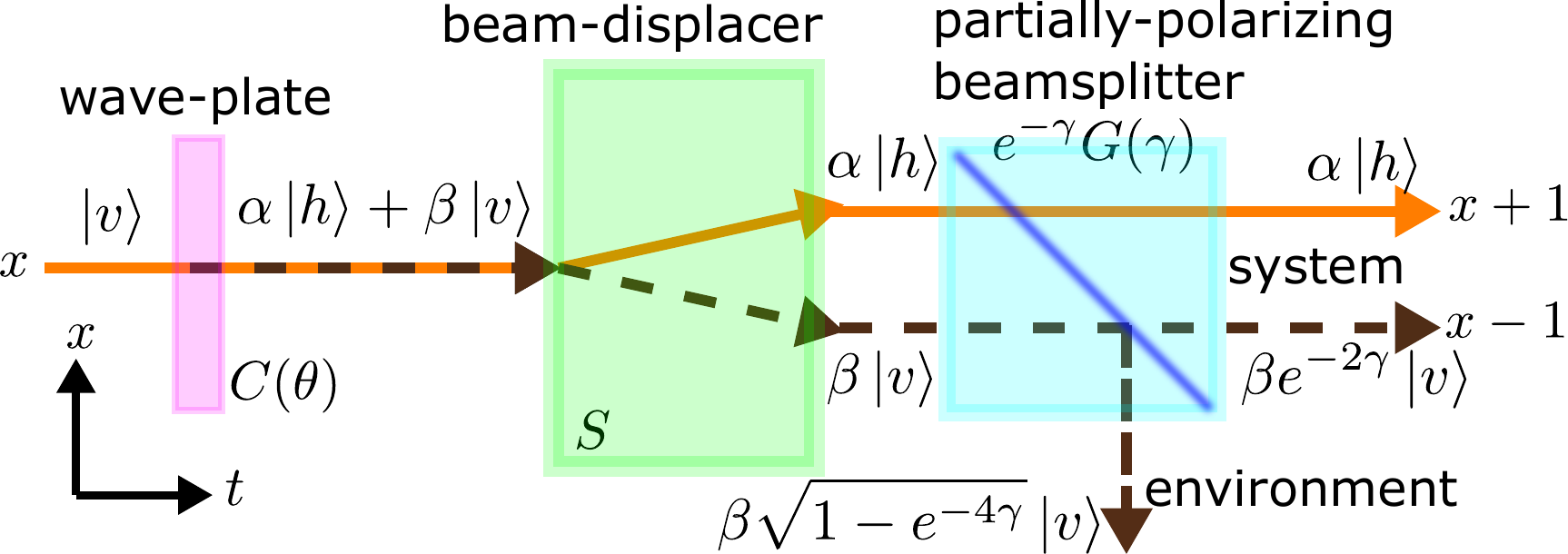}
\caption{Schematic picture for dynamics in which photons pass through optical elements.  
This figure corresponds to a sequence $G(+\gamma)SC(\theta)$ in terms of the matrices in Eq. (2). 
When photons with the polarization $v$ pass through the wave plate, which is described as the shallow purple rectangle, the superposition of $h$ and $v$ is realized with $\alpha=-\sin(\theta)$ and $\beta=\cos(\theta)$. 
The beam displacer, which corresponds to the thick green rectangle, causes the position shift of photons depending on their polarization; photons with $h$ and $v$ respectively go to the right and left. 
When photons enter the partially polarizing beam splitter, which is the blue square, photons with $v$ get into the environment with a probability $1-e^{-4\gamma}$, while photons with $h$ always go straight and remain in the system. 
If we carry out the postselection and focus only on the cases in which all photons are in the system, the creation operators for photons with $v$ acquire the additional factor $e^{-2\gamma}$. 
While unitary dynamics by $C(\theta)$ and $S$ is realized by photons passing through linear optical elements, couplings to environment and postselection are additionally needed for non-unitary dynamics by $G(\gamma)$. }
\label{fig:optical-elements}
\end{center}
\end{figure}

First, we explain the effects of optical elements that do not change the number of photons and lead to unitary dynamics observed in Refs. 
\cite{do2005experimental, broome2010discrete, kitagawa2012observation, zhao2015experimental, cardano2016statistical, xu2019experimental}. 
A wave plate, described by a shallow purple rectangle in Fig. \ref{fig:optical-elements}, changes the polarization of photons. 
We consider the case in which polarizations are altered by rotation matrices parametrized by angles $\theta$ as
\begin{align}
\hat{C}(\theta)=\exp\left[\theta\sum_{x=1}^L(\hat{b}_{x,v}^\dagger\hat{b}_{x,h}-\hat{b}_{x,h}^\dagger\hat{b}_{x,v})\right].
\label{eq:coin_operator}
\end{align}
The unitary transformation of creation operators by $\hat{C}(\theta)$ becomes
\begin{align}
\hspace{-2mm}\left[\begin{array}{c}
\hat{C}(\theta)\hat{b}^\dagger_{x,h}\hat{C}^\dagger(\theta)\\
\hat{C}(\theta)\hat{b}^\dagger_{x,v}\hat{C}^\dagger(\theta)
\end{array}\right]
=\left[\begin{array}{cc}
\cos(\theta)&\sin(\theta)\\
-\sin(\theta)&\cos(\theta)
\end{array}\right]
\left[\begin{array}{c}
\hat{b}^\dagger_{x,h}\\
\hat{b}^\dagger_{x,v}
\end{array}\right].
\label{eq:coin_matrix}
\end{align}
A beam displacer, corresponding to a thick green rectangle, shifts the positions of photons depending on their polarizations, as depicted in Fig. \ref{fig:optical-elements}. 
We consider the case in which photons with the horizontal ($h$) and vertical ($v$) polarizations are shifted to the right and left directions, respectively. 
We can describe such effects by beam displacers utilizing an operator $\hat{S}$ defined as
\begin{align}
\hat{S}=&\exp\left[-i\frac{\pi}{2}\sum_{x=1}^L(\hat{b}_{x,h}^\dagger\hat{b}_{x,v}
+\hat{b}_{x,v}^\dagger\hat{b}_{x,h})\right]\nonumber\\
&\times\exp\left[i\frac{\pi}{2}\sum_{x=1}^L(\hat{b}_{x+1,v}^\dagger\hat{b}_{x,h}+\hat{b}_{x-1,h}^\dagger\hat{b}_{x,v})\right].
\label{eq:shift_operator}
\end{align}
Actually, by the operation of $\hat{S}$, creation operators are transformed as
\begin{align}
\left(\begin{array}{c}
\hat{S}\hat{b}^\dagger_{x,h}\hat{S}^\dagger\\
\hat{S}\hat{b}^\dagger_{x,v}\hat{S}^\dagger
\end{array}\right)
=\left(\begin{array}{c}
\hat{b}^\dagger_{x+1,h}\\
\hat{b}^\dagger_{x-1,v}
\end{array}\right),
\label{eq:shift_matrix}
\end{align}
which corresponds to the position shifts of photons depending on their polarizations. 

Second, we introduce how photons experience non-unitary dynamics, observed in Refs. \cite{xiao2017observation,xiao2018higher,xiao2020non}, on the basis of the combination of the postselection and an optical element referred to as a partially polarizing beam splitter. 
The operation of photons by the partially polarizing beam splitter depends on the polarization. 
For simplicity, we explain effects of partially polarizing beam splitters for $v$, but we can easily understand effects of partially polarizing beam splitters for $h$ by replacing $v$ with $h$. 
If photons pass through the partially polarizing beam splitter for $v$, schematically described by the blue square in Fig. \ref{fig:optical-elements}, photons with $v$ are lost into the environment with a certain probability, while photons with $h$ always remain in the system. 
Such processes can be described by
\begin{align}
\hat{P}_v(\zeta)=\exp\left[\zeta\sum_{x=1}^L
(\hat{b}_{\text{s},x,v}^\dagger
\hat{b}_{\text{e},x,v}
-\hat{b}_{\text{e},x,v}^\dagger
\hat{b}_{\text{s},x,v})\right]
\label{eq:loss_operator_total}
\end{align}
where the subscripts s and e denote the system and environment, respectively. 
For photons with $v$ in the system, $\hat{P}_v(\zeta)$ gives an outcome similar to that of $\hat{C}(\theta)$, while the bases are mixed for s and e instead of $h$ and $v$. 
If $\hat{P}_v(\zeta)$ acts on a state $\ket{\psi_\mathrm{in}}=\prod_{x=x_1}^{x_n}[c_x\hat{b}^\dagger_{\text{s},x,h}
+d_x\hat{b}^\dagger_{\text{s},x,v}]\ket{\mathrm{s}0}\otimes\ket{\mathrm{e}0}$ with $c_x$ and $d_x$ being complex coefficients, the output becomes
\begin{align}
&\ket{\psi_\mathrm{out}}=\hat{P}_v(\zeta)\ket{\psi_\mathrm{in}}\nonumber\\
&=\prod_{x=x_1}^{x_n}[c_x\hat{b}^\dagger_{\text{s},x,h}
+d_x\{\cos(\zeta)\hat{b}^\dagger_{\text{s},x,v}
-\sin(\zeta)\hat{b}^\dagger_{\text{e},x,v}\}]
\ket{\mathrm{s}0}\otimes\ket{\mathrm{e}0},
\label{eq:loss_state}
\end{align}
where $\ket{\mathrm{s}0}$ and $\ket{\mathrm{e}0}$ are the vacuum for the system and environment, respectively. 
Note that $\ket{\mathrm{s}0}$ in this section corresponds to $\ket{0}$ in the main text. 
The output state in Eq. (\ref{eq:loss_state}) includes states where photons are present in the environment, while the input state  $\ket{\psi_\mathrm{in}}$ only includes states where photons are  absent in the environment and all photons exist in the system. 
We carry out the postselection for the state $\ket{\mathrm{e}0}$. 
This corresponds to focusing only on cases in which all photons remain in the system. 
After the postselection, the state of the system becomes
\begin{align}
\ket{\psi_\text{selected}}
&\propto\left\langle\mathrm{e}0|\psi_\mathrm{out}\right\rangle
\nonumber\\
&=\prod_{x=x_1}^{x_n}[c_x\hat{b}^\dagger_{\mathrm{s},x,h}
+e^{-2\gamma}d_x\hat{b}^\dagger_{\mathrm{s},x,v}]\ket{\mathrm{s}0},
\label{eq:post-selection}
\end{align}
where we put $e^{-2\gamma}=\cos(\zeta)$ with introducing a real and positive quantity $\gamma$. 
Therefore, by the partially polarizing beam splitter and the postselection to states with no photon in the environment, the creation operators are transformed as
\begin{align}
&\hat{P}_v[\zeta=\cos^{-1}(e^{-2\gamma})]\,\&\,
\text{postselection}:\nonumber\\
&\left(\begin{array}{c}
\hat{b}^\dagger_{x,h}\\
\hat{b}^\dagger_{x,v}
\end{array}\right)\rightarrow
e^{-\gamma}\left(\begin{array}{cc}
e^{+\gamma}&0\\
0&e^{-\gamma}
\end{array}\right)\left(\begin{array}{c}
\hat{b}^\dagger_{x,h}\\
\hat{b}^\dagger_{x,v}
\end{array}\right).
\label{eq:loss_matrix}
\end{align}
In the main text, we ignore the overall factor $e^{-\gamma}$ since it has no huge effect on dynamics, while it decreases the success probability of postselected dynamics.
Thus, we introduce the matrix $\exp(\gamma\sigma_3)$, which includes the effective amplification of photons with $h$ and is useful for analyzing $\mathcal{PT}$ symmetry breaking. 
This corresponds to the non-unitary matrix $G(\gamma)$ in the main text. 
Note that $x_p=x_q$ with $p \neq q$ is allowed for  $\prod_{x=x_1}^{x_n}$ in Eqs. (\ref{eq:loss_state}) and (\ref{eq:post-selection}). 
Thus, the above discussion can be applied to cases where several photons exist at a position $x$.

\section{Transition between ballistic and diffusive dynamics due to $\mathcal{PT}$ symmetry breaking}
\label{sec:ballistic-diffusive}
We explain that $\mathcal{PT}$ symmetry breaking causes the transition from ballistic dynamics into diffusive dynamics for a single photon. 
As revealed in the next section, the ballistic-diffusive transition profoundly affects the distinguishability transition in short-time dynamics. 

Calculating the moments of $x$, we derive the ballistic-diffusive transition caused by $\mathcal{PT}$ symmetry breaking. 
With the initial state $\ket{\psi(t=0)}=\ket{x^\mathrm{in}=0}\otimes\ket{\sigma^\mathrm{in}}$, the $m$th-order moment can be written as 
\begin{align}
    &\left<x^m(t)\right>=\frac{(-i)^m}{N(t)}\int_{-\pi}^\pi dk
    \bra{\sigma^\mathrm{in}}[\tilde{V}^\dagger(k)]^t\frac{d^m}{dk^m}
    \tilde{V}^t(k)\ket{\sigma^\mathrm{in}}\nonumber\\
    &\simeq\frac{(-i)^m}{N(t)}\int_{-\pi}^\pi dk\sum_sf_s(k)
    e^{i\varepsilon_s^\ast(k)t}
    \frac{d^me^{-i\varepsilon_s(k)t}}{dk^m}
    \label{eq:moment}
\end{align}
where $f_s(k)=|\left<\phi_s^\mathrm{L}(k)|\sigma^\mathrm{in}\right>|^2 \left<\phi_s^\mathrm{R}(k)|\phi_s^\mathrm{R}(k)\right>$ and $N(t)$ is the normalization factor
\begin{align}
    N(t)=\int_{-\pi}^\pi dk
    \bra{\sigma^\mathrm{in}}[\tilde{V}^\dagger(k)]^t
    \tilde{V}^t(k)\ket{\sigma^\mathrm{in}}.
    \label{eq:normalization-factor}
\end{align}
Here, we have used the eigenvalue decomposition of $\tilde{V}^t(k)$
\begin{align}
    \tilde{V}^t(k)&=\exp[-i\tilde{H}(k)t]\nonumber\\
    &=\sum_s e^{-i\varepsilon_s(k)t}
        \ket{\phi_s^\mathrm{R}(k)}\bra{\phi_s^\mathrm{L}(k)}
    \label{eq:t-step}
\end{align}
with $s=\pm$. 
In deriving the second line of Eq. (\ref{eq:moment}), we focus on the leading terms for large $t$, where the sign for the integrant is independent of $k$. 
This is because the other terms should be negligibly small due to the cancellation caused by the oscillation of the integrant as we sweep the Brillouin zone.

\begin{figure}[tbp]
\begin{center}
\includegraphics[width=\columnwidth]{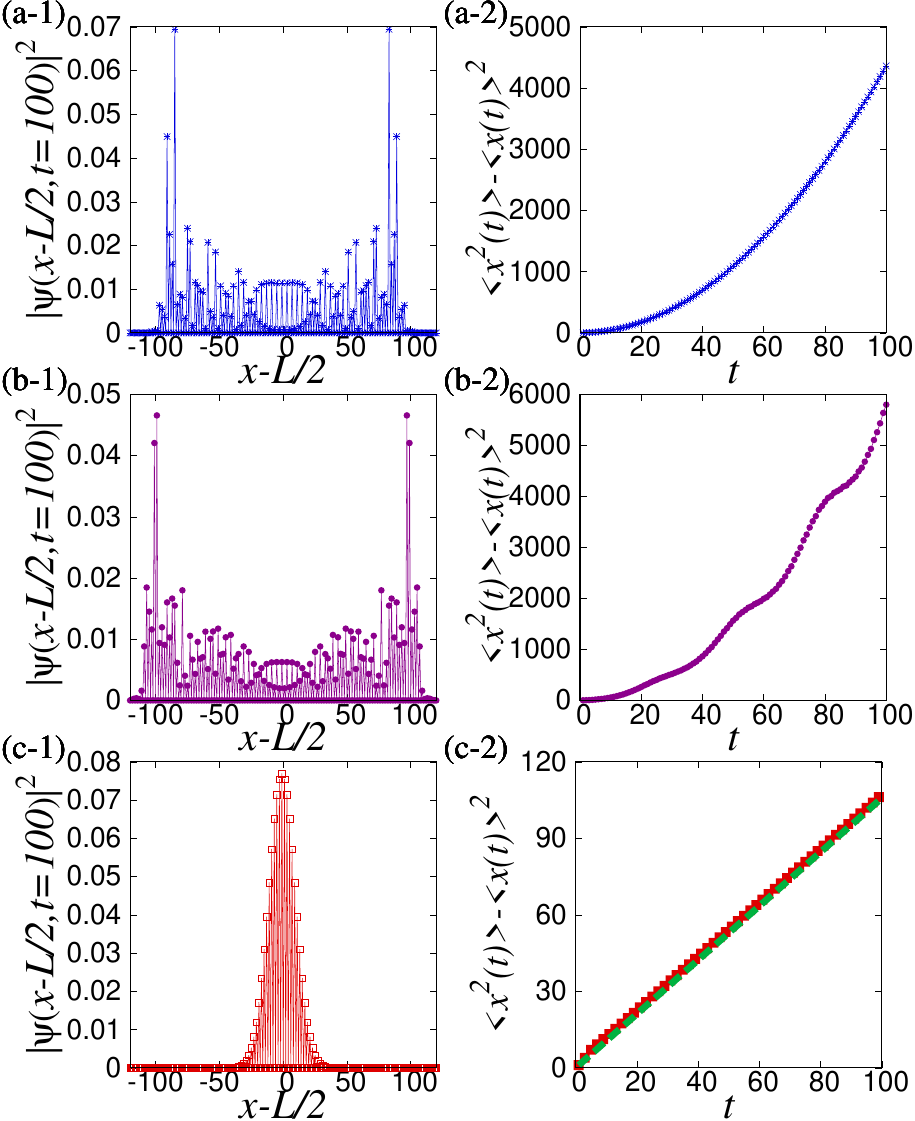}
\caption{Single-photon dynamics of quantum walks where $\theta_1=0.65\pi,\,\theta_2=0.25\pi$, and  $\ket{\psi(t=0)}=\ket{x^\mathrm{in}=L/2}\otimes\ket{\sigma^\mathrm{in}}$ with $\ket{\sigma^\mathrm{in}}=(\ket{h}+i\ket{v})/\sqrt{2}$. 
The parameters for the dissipation are (a) $e^\gamma=1$, (b) $e^\gamma=1.2$, and (c) $e^\gamma=1.5$. 
In the left column, the single-photon probability distributions at $t=100$ are shown. 
In the right column, the variance of $x$ is displayed as a function of time. 
In the (b) $\mathcal{PT}$-symmetric and (c) $\mathcal{PT}$-broken phases, dynamics becomes ballistic and diffusive, respectively. 
In (c-2), the analytically derived variance $Dt/2$ described by the green broken line agrees well with the numerical result described by the red squares. }
\label{fig:probability_deviation}
\end{center}
\end{figure}

First, we show the ballistic behavior in the  $\mathcal{PT}$-symmetric phase by calculating the variance of $x$. 
In the $\mathcal{PT}$-symmetric phase, the dominant term in the variance reads
\begin{align}
    \left<x^2(t)\right>-\left<x(t)\right>^2
    \simeq t^2\frac{\int_{-\pi}^\pi dk\sum_s
    \left[\frac{d\varepsilon_s(k)}{dk}\right]^2f_s(k)}
    {\int_{-\pi}^\pi dk\sum_s f_s(k)}.
    \label{eq:deviation_preserved-ballistic}
\end{align} 
Equation (\ref{eq:deviation_preserved-ballistic}) indicates that dynamics is ballistic, since the coefficient of $t^2$ is always positive. 
Actually, as shown in Fig. \ref{fig:probability_deviation} (b), single-photon dynamics exhibits the ballistic behavior, similar to the unitary case in Fig. \ref{fig:probability_deviation} (a).

Second, we show that dynamics becomes diffusive in the $\mathcal{PT}$-broken phase, i.e., the single-photon distribution asymptotically becomes the Gaussian distribution. 
To see this, we focus only on $\ket{\phi_-^\mathrm{L/R}(k)}$ around $k=0$ since these growing states become dominant over the other states. 
Since $\mathrm{Re}[\varepsilon_-(k)]=0$ around $k=0$ and $\frac{d}{dk}\varepsilon_-(k)=0$ at $k=0$, as demonstrated in Fig. \ref{fig:dispersion} (c), we can approximate the dispersion relation of the quasi-energy as
\begin{align}
    \varepsilon_-(k)\simeq
    \varepsilon_-(k=0)-i\frac{D}{2}k^2
    \label{eq:dispersion_quadratic}
\end{align}
where the coefficient of $k^2$ becomes $D=\left|\frac{d^2}{dk^2}\varepsilon_-(k)\right|_{k=0}=|4\cos(\theta_1)\cos(\theta_2)/\sqrt{d^2(k=0)-1}|$. 
We note that such an approximation is applicable in a wide range of $\mathcal{PT}$-symmetric non-Hermitian systems with translation invariance. 
This is because such systems exhibit dispersion relations similar to that in Fig. \ref{fig:dispersion} (c) if some of the eigenstates break $\mathcal{PT}$ symmetry. 
Therefore, the diffusive behavior which we explain below can be observed in a wide range of $\mathcal{PT}$-symmetric systems. 

On the basis of Eqs. (\ref{eq:moment}) and (\ref{eq:dispersion_quadratic}), the $m$th moment of $x$ in the $\mathcal{PT}$-broken phase becomes 
\begin{align}
    \left<x^m(t)\right>&\simeq
    \frac{(-i)^m\int_{-\infty}^{+\infty} dke^{-\frac{Dk^2}{2}t}\frac{d^m}{dk^m}e^{-\frac{Dk^2}{2}t}}{\int_{-\infty}^{+\infty} dke^{-Dk^2t}}\nonumber\\
    &=\frac{i^m\int_{-\infty}^{+\infty}dke^{-Dk^2t}\left(\frac{Dt}{2}\right)^\frac{m}{2}
    H_m\left(\sqrt{\frac{Dt}{2}}k\right)}
    {\int_{-\infty}^{+\infty} dke^{-Dk^2t}},
    \label{eq:moment_broken}
\end{align}
where $H_m\left(\sqrt{\frac{Dt}{2}}k\right)$ with $m=0,1,2,\cdots$ are the Hermite polynomials. 
In Eq. (\ref{eq:moment_broken}), we have approximated $f_-(k)$ as a constant because $k$-dependent terms in $f_-(k)$ lead to higher-order terms of $1/\sqrt{Dt}$, which are negligible for large $t$.  
We have also extended the range of integration to $(-\infty,+\infty)$ since large-$k$ terms have a low growth rate compared to terms with $k\simeq0$ and thus are negligible. 
Utilizing Eq. (\ref{eq:moment_broken}) and the generating function of the Hermite polynomials $\sum_{m=0}^\infty H_m(y)\frac{z^m}{m!}=\exp(2yz-z^2)$, 
we can obtain the moment generating function of $x$,
\begin{align}
    \left<\exp[\xi x(t)]\right>&\simeq
    \frac{\int_{-\infty}^{+\infty}dk\exp
    \left(-Dk^2t+\frac{Dt}{2}\xi^2+i\xi Dtk\right)}{\int_{-\infty}^{+\infty} dk\exp(-Dk^2t)}\nonumber\\
    &=\exp\left[\frac{\xi^2}{2}\left(
    \sqrt{\frac{Dt}{2}}\right)^2\right].
    \label{eq:moment_generating-function}
\end{align}
The right-hand side of Eq. (\ref{eq:moment_generating-function}) corresponds to the moment generating function of the Gaussian distribution, which means that dynamics exhibits asymptotically diffusive behavior with the standard deviation of $x$ being $\sqrt{Dt/2}$ in the $\mathcal{PT}$-broken phase. 
The green broken line in Fig. \ref{fig:probability_deviation} (c-2) shows $Dt/2$, which indicates that the analytical result agrees well with the numerical result. 

\begin{figure}[tbp]
\begin{center}
\includegraphics[width=\columnwidth]{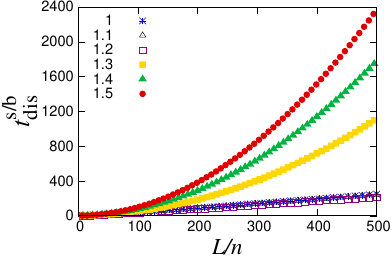}
\caption{System-size dependence of $t_\mathrm{dis}^\mathrm{s}$ and $t_\mathrm{dis}^\mathrm{b}$ with $\theta_1=0.65\pi$ and $\theta_2=0.25\pi$ when two photons $n=2$ are initially located at $x_1^\mathrm{in}=L/2$ with $\sigma_1^\mathrm{in}=h$ and $x_2^\mathrm{in}=L$ with $\sigma_2^\mathrm{in}=v$. 
Each symbol corresponds to a different value of $e^\gamma$. 
The threshold time $t_\mathrm{dis}^\mathrm{s/b}$ in this figure is the smallest time at which $\eta(t)>\delta$ is satisfied with $\delta=10^{-10}$. 
In the $\mathcal{PT}$-symmetric and $\mathcal{PT}$-broken phases, the transition times are linear and quadratic functions of the system size, respectively. }
\label{fig:size-dependence_both}
\end{center}
\end{figure}

\section{Evaluation of the transition time for the short-time distinguishability transition}
\label{sec:evaluation_td}
We evaluate $t_\mathrm{dis}^\mathrm{s/b}$, which is the threshold time for the short-time dynamical distinguishability transition, through single-photon dynamics. 
Figure \ref{fig:size-dependence_both} shows numerically obtained transition times in both $\mathcal{PT}$-symmetric and $\mathcal{PT}$-broken phases, when two photons are initially put at $x_1^\mathrm{in}=L/2$ with $\sigma_1^\mathrm{in}=h$ and $x_2^\mathrm{in}=L$ with $\sigma_2^\mathrm{in}=v$.
In the $\mathcal{PT}$-symmetric phase with  $e^\gamma<e^{\gamma_\mathcal{PT}}\simeq1.22$, $t_\mathrm{dis}^\mathrm{s}$ is proportional to the system size $L$. 
In unitary dynamics, it is known that the threshold time is proportional to the system size, which can be evaluated by the Lieb-Robinson bound \cite{deshpande2018dynamical}.   
In that sense, the behavior in the $\mathcal{PT}$-symmetric phase retains the structure found in unitary dynamics despite gain/loss terms. 
We also note that $t_\mathrm{dis}^\mathrm{s}$ is almost independent of  $\gamma$ and close to the threshold time for the unitary case, which can be seen from Fig. \ref{fig:size-dependence_both}. 
In contrast, in the $\mathcal{PT}$-broken phase, single-photon dynamics becomes diffusive as explained in the previous section. 
Then, $t_\mathrm{dis}^\mathrm{b}$ is proportional to $L^2$ as shown in Fig. \ref{fig:size-dependence_both}, which leads to $t_\mathrm{dis}^\mathrm{b} \gg t_\mathrm{dis}^\mathrm{s}$ in large systems. 

\begin{figure}[tbp]
\begin{center}
\includegraphics[width=\columnwidth]{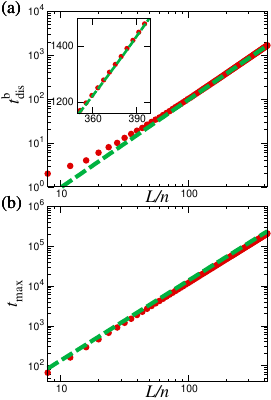}
\caption{System-size dependence of the threshold times at which dynamical distinguishability transitions occur in (a) short-time and (b) long-time scales when $\mathcal{PT}$ symmetry is broken.  
In (a), red circles represent the numerically obtained smallest times at which $\eta(t)>\delta$ is satisfied. 
The green broken line shows  $L^2/2Dn^2|\log(\delta/4n)|$ in the right-hand side of Eq. (\ref{eq:td}), which confirms the analytically derived inequality. 
In (b), red circles represent the threshold times for the long-time distinguishability transition at which $\eta(t)<\delta$ is satisfied for the first time in numerical simulations with the constraint that $t$ is larger than $t_\mathrm{dis}^\mathrm{b}$. 
The green broken line represents $L^2|\log(\delta/32n^2)|/D(2\pi)^2$ in the right hand side of Eq. (\ref{eq:tm}), which verifies the analytical result for sufficiently large $L$. 
In both (a) and (b), rotation angles are $\theta_1=0.65\pi,\,\theta_2=0.25\pi$, the gain-loss parameter is $e^\gamma=1.5$, the number of photons is $n=2$, the initial state is  $\hat{b}_{x=L/2,h}^\dagger\hat{b}_{x=L,v}^\dagger\ket{0}$, and $\delta=10^{-10}$.}
\label{fig:size-dependence_broken}
\end{center}
\end{figure}

To evaluate $t_\mathrm{dis}^\mathrm{b}$ more quantitatively in the $\mathcal{PT}$-broken phase, we approximate the single-photon wave functions by Gaussian distributions on the basis of Eq. (\ref{eq:moment_generating-function}). 
Also, we ignore the polarization $\sigma=h,v$ for simplicity. 
We consider a situation in which $n$ photons are initially put at $x_p^\mathrm{in}=\frac{L}{n}p$ with $p=1,2,\cdots,n$, where $Lp/n$ corresponds to $\mathrm{In}_p$ in the main text. 
In this case, the single-photon wave functions become
\begin{align}
    \psi_p(x,t)\propto
    \exp\left[-\frac{\left(x-\frac{L}{n}p\right)^2}{2Dt}\right].
    \label{eq:gaussian-distribution}
\end{align}
The matrix elements of $W(t)$ become $W_{pq}(t)=\psi_p(x_q^\mathrm{out},t)$.
Here, $x_q^\mathrm{out}$ with $q=1,2,\cdots,n$ is a position at which an output photon is detected, which corresponds to $\mathrm{Out}_q$ in the main text. 
While phases on $\psi_p(x_q^\mathrm{out},t)$ can take various values depending on $x_q^\mathrm{out}$ and $t$, we neglect effects of these phases. 
The ignorance of the phases is consistent with an inequality derived in Eq.~\eqref{eq:inequality_difference} below.
Indeed, it means that we evaluate the $L_1$-distance between distributions of photons and that of distinguishable particles larger than the actual value, not by taking cancellations of numerous terms with various phases into consideration. 
We write the difference between the actual photon distribution and the distribution of distinguishable particles as
\begin{align}
    P({\bm y})-P_\mathrm{dis}({\bm y})
    \simeq\frac{\tilde{P}_\mathrm{dis}({\bm y})+R({\bm y})}{N(t)}
    -\frac{\tilde{P}_\mathrm{dis}({\bm y})}{N_\mathrm{dis}(t)},
    \label{eq:difference}
\end{align}
where ${\bm y}=(x_1^\mathrm{out},x_2^\mathrm{out},\cdots,x_n^\mathrm{out},t)$. 
Through Eq. (4) in the main text, we can understand that the terms in the numerator become
\begin{align}
    \tilde{P}_\mathrm{dis}({\bm y})&=\sum_\omega\prod_{p=1}^n
    |\psi_p({\bm y}_{\omega[p]})|^2,
    \label{eq:Pd}\\
    R({\bm y})&=\sum_{\omega\neq\tau}\prod_{p=1}^n\psi_p({\bm y}_{\omega[p]})\psi_p^*({\bm y}_{\tau[p]}),
    \label{eq:R}
\end{align}
where $\omega$ and $\tau$ in the sum are taken all over permutations, and ${\bm y}_q=(x_q^\mathrm{out},t)$. 
Here, we dropped $\prod_{j=1}^Mn_j^\mathrm{out}!$ assuming that  probabilities where several photons are detected at the same position are negligibly small. 
In Eq. (\ref{eq:difference}), normalization factors $N(t)$ and $N_\mathrm{dis}(t)$ become
\begin{align}
    N(t)&=N_\mathrm{dis}(t)+\sum_{\{x^\mathrm{out}\}}R({\bm y}),
    \label{eq:normalization_actual}\\
    N_\mathrm{dis}(t)&=\sum_{\{x^\mathrm{out}\}}
    \tilde{P}_\mathrm{dis}({\bm y}).
    \label{eq:normalization_distinguishable}
\end{align}
Here, we assume $(L/n)^2 \gg Dt$ for analyzing the short-time distinguishability transition. 
For evaluating $R({\bm y})$, we focus on a product of a specific pair $\psi_p({\bm y}_{\omega[p]})$ and $\psi_{p'}^*({\bm y}_{\tau[p']})$ in $R({\bm y})$, where the permutations for $p$ and $p'$ coincide, i.e., $\omega(p)=\tau(p')=q$. 
Then, if we write $p'=p+\delta p$, we can evaluate $\psi_p(x_q^\mathrm{out},t)\psi_{p+\delta p}^*(x_q^\mathrm{out},t)$ as $\exp\left[-\frac{L^2(\delta p)^2}{4n^2Dt}\right]\left|\tilde{\psi}_{p+\delta p/2}(x_q^\mathrm{out})\right|^2$ with $\tilde{\psi}_{p+\delta p/2}(x)$ being the right-hand side of Eq. (\ref{eq:gaussian-distribution}), where $p+\delta p/2$ is substituted into $p$. 
Therefore, large $\delta p$ leads to higher order terms of $\exp\left(-\frac{L^2}{4n^2Dt}\right)$.  
Thus, due to the constraint $\omega\neq\tau$, leading terms of $R(\{\bm y\})$ become
\begin{align}
    R({\bm y})\simeq\sum_{\omega,p}&
    \prod_{p_1=1}^{p-1}|\psi_{p_1}({\bm y}_{\omega[p_1]})|^2
    \prod_{p_2=p+2}^n|\psi_{p_2}({\bm y}_{\omega[p_2]})|^2
    \nonumber\\&\times
    2\mathrm{Re}\left[\Psi_p({\bm y}_{\omega[p]})
    \Psi_p^*({\bm y}_{\omega[p+1]})\right],
    \label{eq:R_approximation}
\end{align}
where $\Psi_p({\bm y}_{\omega[p]})=\psi_{p}({\bm y}_{\omega[p]})\psi_{p+1}^*({\bm y}_{\omega[p]})$. 
On the basis of Eq. (\ref{eq:R_approximation}), we can evaluate the ratio of $N_\mathrm{dis}(t)$ and $\sum_{\{x^\mathrm{out}\}}R({\bm y})$ as
\begin{align}
    \frac{\sum_{\{x^\mathrm{out}\}}R({\bm y})}
    {\sum_{\{x^\mathrm{out}\}}\tilde{P}_\mathrm{dis}({\bm y})}
    \simeq 2n\exp\left(-\frac{L^2}{2n^2Dt}\right).
    \label{eq:ratio_PR}
\end{align}
Here, the factor $n$ on the right-hand side originates from the number of cases where we choose a pair $p$ and $p+1$. 
Thus, through the Taylor expansion of $1/N(t)$, the difference between the probability distribution of photons and that of the distinguishable particles becomes
\begin{align}
    P({\bm y})-P_\mathrm{dis}({\bm y})
    \simeq\frac{R({\bm y})
    -\tilde{P}_\mathrm{dis}({\bm y})
    2ne^{-\frac{L^2}{2n^2Dt}}}{N_\mathrm{dis}(t)}
    \label{eq:difference_evaluation}
\end{align}
for $(L/n)^2 \gg Dt$. Equation (\ref{eq:difference_evaluation}) results in an inequality of the $L_1$-distance between $P({\bm y})$ and $P_\mathrm{dis}({\bm y})$,
\begin{align}
    &\eta(t)
    =\sum_{\{x_q^\mathrm{out}\}}\left|P({\bm y})-P_\mathrm{dis}({\bm y})\right|
    \nonumber\\
    &\leq\sum_{\{x_q^\mathrm{out}\}}
    \left|\frac{R({\bm y})}{N_\mathrm{dis}(t)}\right|
    +2n\exp\left(-\frac{L^2}{2n^2Dt}\right)
    \sum_{\{x_q^\mathrm{out}\}}
    \left|\frac{\tilde{P}_\mathrm{dis}({\bm y})}
    {N_\mathrm{dis}(t)}\right|
    \nonumber\\
    &\simeq4n\exp\left(-\frac{L^2}{2n^2Dt}\right).
    \label{eq:inequality_difference}
\end{align}
Therefore, the actual threshold time $t_\mathrm{dis}^\mathrm{b}$ at which $\eta(t)>\delta$ is satisfied for the first time is larger than the time at which $4n\exp\left(-L^2/2n^2Dt\right)\simeq\delta$ is satisfied. 
This discussion leads to
\begin{align}
    t_\mathrm{dis}^\mathrm{b}
    >\frac{L^2}{2Dn^2|\log(\delta/4n)|}.
    \label{eq:td}
\end{align}
The green broken line in Fig. \ref{fig:size-dependence_broken} (a) represents the right-hand side of Eq. (\ref{eq:td}) with $\delta=10^{-10}$, which is below the numerically obtained $t_\mathrm{dis}^\mathrm{b}$, and thus verifies our inequality. 

\section{Evaluation of the transition time for the long-time distinguishability transition}
\label{sec:evaluation_tm}
We evaluate the threshold time $t_\mathrm{max}$ of the distinguishability transition for long-time dynamics by the gap size of the imaginary part of quasi-energies. 
Since the computability of $P(\{n^\mathrm{in}\},\{n^\mathrm{out}\},t)$ is based on the dominant eigenstate $\ket{\phi_\mathrm{max}^\mathrm{L/R}}$, the gap size $\Delta=|\varepsilon_\mathrm{max}-\varepsilon_a|$ between $\varepsilon_\mathrm{max}=i\log(\lambda_\mathrm{max})$ and $\varepsilon_a=i\log(\lambda_a)$ is important, where $|\lambda_\mathrm{max}|=\max_l|\lambda_l|\equiv|\lambda_{l=l_\mathrm{max}}|$ and $|\lambda_a|=\max_{l (\neq l_\mathrm{max})}|\lambda_l|\equiv|\lambda_{l=l_a}|$. Note that $\lambda_a$ can be degenerate. 
Then, $\Delta^{-1}$ determines the timescale where the second dominant eigenstates $\ket{\phi_a^\mathrm{L/R}}$ become negligible compared to the dominant state $\ket{\phi_\mathrm{max}^\mathrm{L/R}}$. 
In the present model, $\varepsilon_\mathrm{max}=\varepsilon_-(k=0)$ and $\varepsilon_a=\varepsilon_-(k=\pm2\pi/L)$, which results in
\begin{align}
    \Delta=\frac{D}{2}\left(\frac{2\pi}{L}\right)^2
    \label{eq:gap-size}
\end{align}
within the approximation in Eq. (\ref{eq:dispersion_quadratic}). 
We evaluate $\eta(t)$ by taking terms proportional to $e^{- g \Delta t}$ into consideration, where $g=0,1,2$. 
After long time evolution, $U^\mathrm{T}(t)=V^t$ becomes
\begin{align}
    U^\mathrm{T}(t)\simeq \lambda_\mathrm{max}^t\biggl(
    \ket{\phi_\mathrm{max}^\mathrm{R}}
    &\bra{\phi_\mathrm{max}^\mathrm{L}}
    +e^{- \Delta t}\sum_a
    \ket{\phi_a^\mathrm{R}}
    \bra{\phi_a^\mathrm{L}}\nonumber\\
    &+e^{-2 \Delta t}\sum_b
    \ket{\phi_b^\mathrm{R}}
    \bra{\phi_b^\mathrm{L}}\,\biggr),
    \label{eq:evaluation_U}
\end{align}
where the third dominant components with $|\lambda_b|=\max_{l (\neq l_\mathrm{max},l_a)}|\lambda_l|$ are also included. 
Here, the sums for $a$ and $b$ include two terms that respectively correspond to $k=\pm2\pi/L$ and $\pm4\pi/L$ with $s=-$. 
We assume that dependence of $|\left\langle\mathsf{Out}_q|\phi_c^\mathrm{R}\right\rangle\left\langle\phi_c^\mathrm{L}|\mathsf{In}_p\right\rangle/\left\langle\mathsf{Out}_q|\phi_\mathrm{max}^\mathrm{R}\right\rangle|\left\langle\phi_\mathrm{max}^\mathrm{L}|\mathsf{In}_p\right\rangle|$ on $L$ with $c=a,b$ is negligible for arbitrary $p$ and $q$, compared to $e^{- \Delta t}$ and $e^{-2 \Delta t}$.  
The matrix elements of $W(t)$, which determines the photon probability distribution, can be evaluated as
\begin{align}
    W_{pq}^*(t)\simeq &(\lambda_\mathrm{max}^*)^t\left\langle\mathsf{Out}_q|\phi_\mathrm{max}^\mathrm{R}\right\rangle
    \left\langle\phi_\mathrm{max}^\mathrm{L}|\mathsf{In}_p\right\rangle\nonumber\\
    &\left[1+\sum_aR_q^aL_p^ae^{- \Delta t}+\sum_bR_q^bL_p^be^{-2 \Delta t}\right],
    \label{eq:evaluation_W}
\end{align}
where 
$R_q^c=\left\langle\mathsf{Out}_q|\phi_c^\mathrm{R}\right\rangle/\left\langle\mathsf{Out}_q|\phi_\mathrm{max}^\mathrm{R}\right\rangle$ and $L_p^c=\left\langle\phi_c^\mathrm{L}|\mathsf{In}_p\right\rangle/\left\langle\phi_\mathrm{max}^\mathrm{L}|\mathsf{In}_p\right\rangle$. 
The permanent of $W^*(t)$ becomes
\begin{align}
    \mathrm{Per}[W^*(t)]\simeq &(\lambda_\mathrm{max}^*)^{nt}\mathrm{Per}[Z(t)]\nonumber\\
    &\prod_{p=1}^n\left\langle\mathsf{Out}_p|\phi_\mathrm{max}^\mathrm{R}\right\rangle
    \left\langle\phi_\mathrm{max}^\mathrm{L}|\mathsf{In}_p\right\rangle,
    \label{eq:evaluation_permanent-W}
\end{align}
where $Z_{pq}(t)=1+\sum_aR_q^aL_p^ae^{- \Delta t}+\sum_bR_q^bL_p^be^{-2 \Delta t}$. 
Neglecting higher order terms of $e^{- \Delta t}$, we can obtain the permanent of $Z(t)$,
\begin{align}
    \mathrm{Per}[Z(t)]\simeq
    n!+Q_1e^{- \Delta t}+(Q_2+Q_2')e^{-2 \Delta t},
    \label{eq:evaluation_permanent_Z}
\end{align}
where
\begin{align}
    &Q_1=\sum_\omega\sum_{p=1}^n
    \sum_aR_p^aL_{\omega(p)}^a,
    \label{eq:Q1}\\
    &Q_2=\sum_\omega\sum_{p \neq q}
    \sum_{a,\tilde{a}}R_p^aL_{\omega(p)}^a
    R_q^{\tilde{a}}L_{\omega(q)}^{\tilde{a}},
    \label{eq:Q2}\\
    &Q_2'=\sum_\omega\sum_{p=1}^n
    \sum_bR_p^bL_{\omega(p)}^b
    \label{eq:Q2_dash}
\end{align}
with $\omega$ being the set of all possible permutations.
Thus, the probability distribution of photons can be approximated as
\begin{align}
    &P(\{n^\mathrm{in}\},\{n^\mathrm{out}\},t)
    \nonumber\\
    &\simeq\frac{\prod_p|\langle\mathsf{Out}_p|\phi_\mathrm{max}^\mathrm{R}\rangle|^2}
    {N_\mathrm{max}[1+A(t)]\prod_jn_j^\mathrm{out}!}
    \biggl[1+\frac{2\mathrm{Re}(Q_1)}{n!}e^{- \Delta t}
    \nonumber\\&\ \ \ \ \ \ \ \ 
    +\frac{|Q_1|^2}{(n!)^2}e^{-2 \Delta t}
    +\frac{2\mathrm{Re}(Q_2+Q_2')}{n!}e^{-2 \Delta t}\biggr],
    \label{eq:actual-probability_long-time}
\end{align}
where
\begin{align}
    &N_\mathrm{max}=\sum_{\{n^\mathrm{out}\}}\frac{\prod_p|\left\langle\mathsf{Out}_p|\phi_\mathrm{max}^\mathrm{R}\right\rangle|^2}{\prod_jn_j^\mathrm{out}!}
    \label{eq:N_max}\\
    &A(t)=\sum_{\{n^\mathrm{out}\}}\frac{\prod_p|\langle\mathsf{Out}_p|\phi_\mathrm{max}^\mathrm{R}\rangle|^2}{N_\mathrm{max}\prod_jn_j^\mathrm{out}!}
    \biggl[\frac{2\mathrm{Re}(Q_1)}{n!}e^{- \Delta t}
    \nonumber\\&\ \ \ \ \ \ \ \ \ \ \ \ \ \ 
    +\left(\frac{|Q_1|^2}{(n!)^2}+\frac{2\mathrm{Re}[Q_2+Q_2']}{n!}\right)
    e^{-2 \Delta t}\biggr].
    \label{eq:A}
\end{align}
In order to evaluate $\eta(t)$, we also calculate $P_\mathrm{dis}(\{n^\mathrm{out}\},t)$. 
From Eq. (\ref{eq:evaluation_U}), the distribution of distinguishable particles becomes
\begin{align}
    &P_\mathrm{dis}(\{n^\mathrm{in}\},\{n^\mathrm{out}\},t)
    \nonumber\\&
    \simeq\frac{\prod_p|\langle\mathsf{Out}_p|\phi_\mathrm{max}^\mathrm{R}\rangle|^2}
    {N_\mathrm{max}[1+B(t)]\prod_jn_j^\mathrm{out}!}
    \nonumber\\&\ \ \ \  
    \left[1+\frac{2\mathrm{Re}(Q_1)}{n!}e^{- \Delta t}
    +\frac{2\mathrm{Re}(\tilde{Q}_2+Q_2')}{n!}e^{-2 \Delta t}\right],
    \label{eq:distinguishable-probability_long-time}
\end{align}
where
\begin{align}
    &\tilde{Q}_2=\sum_\omega\sum_{p,q}
    \sum_{a,\tilde{a}}R_p^aL_{\omega(p)}^a
    [R_q^{\tilde{a}}L_{\omega(q)}^{\tilde{a}}]^\ast,
    \label{eq:Q2}\\
    &B(t)=\sum_{\{n^\mathrm{out}\}}\frac{\prod_p|\langle\mathsf{Out}_p|\phi_\mathrm{max}^\mathrm{R}\rangle|^2}{N_\mathrm{max}\prod_jn_j^\mathrm{out}!}
    \biggl[\frac{2\mathrm{Re}(Q_1)}{n!}e^{- \Delta t}
    \nonumber\\&\ \ \ \ \ \ \ \ \ \ \ \ \ \ \ \ \ 
    +\frac{2\mathrm{Re}(\tilde{Q}_2+Q_2')}{n!}e^{-2 \Delta t}\biggr].
    \label{eq:B}
\end{align}
Through the Taylor expansion $[1+C(t)]^{-1}\simeq1-C(t)+C^2(t)/2$ with $C(t)=A(t),B(t)\ll1$, the difference between $P(\{n^\mathrm{out}\},t)$ and $P_\mathrm{dis}(\{n^\mathrm{out}\},t)$ becomes 
\begin{align}
    &\eta(t)=\sum_{\{n^\mathrm{out}\}}|P(\{n^\mathrm{out}\},t)-P_\mathrm{dis}(\{n^\mathrm{out}\},t)|\nonumber\\&
    \simeq\sum_{\{n^\mathrm{out}\}}\frac{\prod_p|\left\langle\mathsf{Out}_p|\phi_\mathrm{max}^\mathrm{R}\right\rangle|^2}{N_\mathrm{max}n!\prod_jn_j^\mathrm{out}!}e^{-2 \Delta t}\nonumber\\&\ \ \ \ \ \ \ \ \ \ \ \  \left|2\mathrm{Re}(Q_2-\tilde{Q}_2)-E+\frac{|Q_1|^2}{n!}\right|\nonumber\\&\leq
    e^{-2 \Delta t}\sum_{\{n^\mathrm{out}\}}\frac{\prod_p|\left\langle\mathsf{Out}_p|\phi_\mathrm{max}^\mathrm{R}\right\rangle|^2}{N_\mathrm{max}\prod_jn_j^\mathrm{out}!}\nonumber\\
    &\ \ \ \ \ \ \ \ \ \ \ \ \ \ \ \ \ \ \ \ \  \left[\frac{4}{n!}(|Q_2|+|\tilde{Q}_2|)+\frac{2|Q_1|^2}{(n!)^2}\right],
    \label{eq:evaluation_difference}
\end{align}
where
\begin{align}
    E=\sum_{\{n^\mathrm{out}\}}&    \frac{\prod_p|\langle\mathsf{Out}_p|\phi_\mathrm{max}^\mathrm{R}\rangle|^2}{N_\mathrm{max}\prod_jn_j^\mathrm{out}!}\nonumber\\
    &\ \ \left[2\mathrm{Re}(Q_2-\tilde{Q}_2)
    +\frac{|Q_1|^2}{n!}\right].
    \label{eq:E}
\end{align}
Here, we put an assumption $|\left\langle\mathsf{Out}_p   |\phi_\mathrm{max}^\mathrm{R}\right\rangle|\simeq1/\sqrt{M}$, $|R_p^c|\simeq1$, and $|L_p^c|\simeq1$ for arbitrary $p$, which is valid in translational invariant systems. 
Then the $L_1$-norm of $P(\{n^\mathrm{in}\},\{n^\mathrm{out}\},t)-P_\mathrm{dis}(\{n^\mathrm{in}\},\{n^\mathrm{out}\},t)$ can be evaluated as
\begin{align}
    \eta(t)\leq32n^2e^{-2 \Delta t}.
    \label{eq:evaluation_norm}
\end{align}
Equation (\ref{eq:evaluation_norm}) results in
\begin{align}
    t_\mathrm{max}\leq-\frac{L^2}{D(2\pi)^2}
    \log\left[\frac{\delta}{32n^2}\right],
    \label{eq:tm}
\end{align}
where $t_\mathrm{max}$ is the time at which $\eta(t)$ becomes smaller than $\delta$ for the first time under the constraint $t>t_\mathrm{dis}^\mathrm{s/b}$. 
Figure \ref{fig:size-dependence_broken} (b) shows that the derived inequality is indeed satisfied.

\bibliographystyle{apsrev4-2}
\bibliography{reference.bib}

\end{document}